\pdfoutput=1

\documentclass[11pt,a4paper]{iopart}

\usepackage[nooneline]{subfigure}
\usepackage{dsfont}
\usepackage{amsopn}

\usepackage[pdftex]{graphicx}
\usepackage{bm,bbm}

\expandafter\let\csname equation*\endcsname\relax
\expandafter\let\csname endequation*\endcsname\relax

\usepackage{iopams}
\usepackage{amsmath,amssymb}

\newcommand{\bra}[1]{\langle #1 |}
\newcommand{\ket}[1]{| #1 \rangle}

\newcommand{\1}{\mathds{1}}
\newcommand{\ii}{\mathrm{i}}
\newcounter{shadowtheorems}

\newtheorem{proposition}[shadowtheorems]{Proposition}

\newcommand{\RP}[1]{{\mathbbm R}\mathbf{P}^{#1}}
\newcommand{\CP}[1]{{\mathbbm C }\mathbf{P}^{#1}}
\newcommand{\Sphere}[1]{\mathbf{S}^{#1}}

\newcommand{\stackidx}[4]{
\substack{
#1 #2 \\
#3 #4}
}

\begin{document}

\title{Restricted numerical shadow and geometry of quantum entanglement}

\author{
Zbigniew Pucha{\l}a$^1$ 
Jaros{\l}aw A. Miszczak$^1$,  
Piotr Gawron$^1$,
Charles~F.~Dunkl$^2$, 
John~A.~Holbrook$^3$, 
and 
Karol~{\.Z}yczkowski$^{4,5}$}

\address{$^1$ Institute of Theoretical and Applied Informatics, Polish Academy
of Sciences, Ba{\l}tycka 5, 44-100 Gliwice, Poland} 

\address{$^2$ Department of Mathematics, University of Virginia, 
Charlottesville, VA 22904---4137, USA} 

\address{$^3$ Department of Mathematics and Statistics, University of Guelph,
Guelph, Ontario, N1G 2W1, Canada}

\address{$^4$ Institute of Physics,  Jagiellonian University, 
 Reymonta 4, 30-059 Krak{\'o}w, Poland} 

\address{$^5$ Center for Theoretical Physics, Polish Academy of Sciences, 
         Aleja Lotnik{\'o}w 32/44, 02-668 Warszawa, Poland}

\ead{z.puchala@iitis.pl \quad miszczak@iitis.pl \quad gawron@iitis.pl \quad
cfd5z@virginia.edu \quad jholbroo@uoguelph.ca \quad karol@tatry.if.uj.edu.pl}


\begin{abstract}
The restricted numerical range $W_R(A)$ of an operator $A$ acting on a
$D$-dimensional Hilbert space is defined as a set of all possible expectation
values of this operator among pure states which belong to a certain subset $R$
of the of set of pure quantum states of dimension $D$. One considers for
instance the set of real states, or in the case of composite spaces, the set of
product states and the set of maximally entangled states. Combining the operator
theory with a probabilistic approach we introduce the restricted numerical
shadow of $A$ -- a normalized probability distribution on the complex plane
supported in $W_R(A)$. Its value at point $z \in {\mathbbm C}$ is equal to the
probability that the expectation value $\langle \psi|A|\psi\rangle$ is equal to
$z$, where $|\psi\rangle$ represents a random quantum state in subset $R$
distributed according to the natural measure on this set, induced by the
unitarily invariant Fubini--Study measure. Studying restricted shadows of
operators of a fixed size $D=N_A N_B$ we analyse the geometry of sets of
separable and maximally entangled states of the $N_A \times N_B$ composite
quantum system. Investigating trajectories formed by evolving quantum states
projected into the plane of the shadow we study the dynamics of quantum
entanglement. A similar analysis extended for operators on $D=2^3$ dimensional
Hilbert space allows us to investigate the structure of the orbits of $GHZ$ and
$W$ quantum states of a three--qubit system.
\end{abstract}

\maketitle

\section{Introduction}
Recent studies on quantum entanglement, a crucial resource in the theory of
quantum information processing, contributed a lot to our understanding of this
deeply non-classical phenomenon (see \emph{e.g.} \cite{HHHH09} and references
therein). In particular, some progress has been achieved in elucidating the
geometry of quantum entanglement \cite{KZ01,MD01,VDM02,Le04,BK09,AK09}, but
several questions concerning this topic remain still open
\cite{BZ06,BGH07,SHK11}. It was also suggested that a geometric approach is
useful to classify and quantify quantum entanglement \cite{BZ06, Ve07,Pe08}.

The phenomenon of quantum entanglement, non-classical correlations between
individual subsystems, may arise in composite physical systems. Consider then
the simplest composite system, which consists of two parts and can be described
in a Hilbert space with a tensor product structure, ${\cal H}_D={\cal
H}_A\otimes {\cal H}_B= {\cal H}_{N_A}\otimes {\cal H}_{N_B}$. Any product pure
state, $\ket{\psi} = \ket{\phi_A} \otimes \ket{\phi_B}$, is called
\emph{separable}, while all other pure states are called \emph{entangled}. The
set $\cal S$ of all separable pure states has the structure of the Cartesian
product of the complex projective spaces \cite{BZ06}, ${\cal S}=\CP{{N_A}-1}
 \times \CP{{N_B}-1}$.

Among entangled pure states of a bipartite system one distinguishes the set
$\cal E$ of \emph{maximally entangled} states, such that the partial trace of
the corresponding projector is proportional to the identity matrix. The set
$\cal E$ contains the generalized Bell state, $|\psi_+\rangle=\frac{1}{\sqrt{N}}
\sum_i |i,i\rangle$, and all states obtained from it by a local unitary
transformation, $U_A\otimes U_B |\psi_+\rangle \in {\cal E}$. The set of
maximally entangled states is thus equivalent to $\mathrm{U}(N)/\mathrm{U}(1)$. 
The structure of the set $\cal E$ and other sets
of locally equivalent entangled states of a bipartite system was studied in
\cite{SZK02}.

In this work we propose to analyse the geometry of the set of entangled and
separable quantum states using the algebraic concepts of the numerical range and
numerical shadow of an operator.
For any operator $A$ acting on the complex Hilbert space ${\cal H}_D$ one
defines its \emph{numerical range} \cite{HJ2,GR97} as a subset of the complex
plane which contains expectation values of $A$ among arbitrary normalized pure
states,
\begin{equation}
W(A)=\{ z: z=\langle \psi | A| \psi \rangle, |\psi\rangle \in {\cal H}_D, 
 \  \langle \psi| \psi \rangle =1\}.
\label{range1}
\end{equation}
Due to the classical Toeplitz-Hausdorff theorem the set $W(A)$ is convex -- see
e.g. \cite{Gu04}. The differential topology and projection aspects of numerical
range were investigated in \cite{JAG98,He10} 

The standard notion of numerical range,
 often used in the theory of quantum information \cite{SHDHG08,KPLRS09,GPMSCZ09},
can be generalized in several ways \cite{MW80,BLP91,DHKSH08}.
For instance, for an operator acting on a composite Hilbert space 
one defines the \emph{product} numerical range \cite{PGMSCZ10}
(also called \emph{local} numerical range \cite{DHKSH08})
and a more general class of numerical ranges \emph{restricted}
 to a specific class of states \cite{GPMSCZ09},
\begin{equation}
W_R(A)=\{ z: z=\langle \psi | A| \psi \rangle, \   |\psi\rangle
 \in R \subset  \Omega_D, 
 \  \langle \psi| \psi \rangle =1\}.
\label{rest_range}
\end{equation}
Here $ R \subset \Omega_D$ denotes a selected subset of the set of pure quantum
states of a given size $D$. For instance, one may consider the set of all real
states, or, in the case of composite spaces, the set of complex product states
or the set of real maximally entangled states.

For any operator $A$ acting on ${\cal H}_D$
one defines a probability density 
$P_A(z)$ on the complex plane \cite{DGHPZ11,GS10},
 supported in the numerical range $W(A)$,
\begin{equation}
P_A(z) \ := \ \int_{\Omega_D} {\rm d} \mu(\psi) \
 \delta\Bigl( z-\langle \psi|A|\psi\rangle\Bigr)  \ .
\label{shadow}  
\end{equation}
Here $\mu(\psi)$ denotes the unique unitarily invariant measure on the set
$\Omega_D$ of $D$--dimensional quantum pure states, also called Fubini-Study
(FS) measure. In other words the shadow of $A$ at a given point $z \in {\mathbbm
C}$ characterizes the likelihood that the expectation of $A$ among a random pure
states is equal to $z$. Sometimes it is convenient to treat the numerical shadow
as a probability measure $\mu_A$, on a complex plane. For any measurable set $E$
it reads $\mu_A(E) = \mu(\{\psi : \bra{\psi}A\ket{\psi} \in E\})$.

If the operator $A$ is normal, $AA^{\dagger}=A^{\dagger}A$, its shadow $P_A(z)$
can be interpreted as a projection of the set of classical states -- the
$D$--dimensional regular simplex of probability distributions -- into a
two-plane \cite{DGHMPZ11}. In the more general case of a non--normal $A$ its
shadow can be associated with a probability distribution obtained by projecting
the set $\Omega_D$ of quantum pure states of size $D$ into a plane. Thus
choosing a matrix $A$ to be analysed one fixes the relative position of the set
$\Omega_D$ and determines the direction, along which it is projection on the
plane. Hence investigating all possible shadows of various operators of a given
size $D$ one gathers information about the structure of the set $\Omega_D$.

Combining the ideas of the restricted numerical range  (\ref{rest_range})
with the numerical shadow (\ref{shadow})
one is led in a natural way to the definition of a
\emph{restricted numerical shadow},
\begin{equation}
P^R_A(z) \ := \ \int_{R} {\rm d} \mu(\psi) \
 \delta\Bigl( z-\langle \psi|A|\psi\rangle\Bigr)  \ ,
\label{rshadow}  
\end{equation}
where $ R$ is the selected subset of ${\Omega}_D$. In particular, choosing the
appropriate subsets of the set of pure states we define the numerical shadow
restricted to  \emph{real} states or $\mathrm{SU}(2)$--coherent states. In the case of a
composite Hilbert space one defines the shadow with respect to \emph{separable}
states or \emph{maximally entangled} states. The restrictions can be combined so
one can consider the shadow restricted e.g. to real separable states, or in the
case of operators acting on $8$ dimensional space,  which describes a
three--qubit system, one may study the shadow with respect to real $GHZ$ states
or complex $W$--states. It will be convenient to use simplified terms, so for
brevity we will slightly abuse the notation and write about 'separable shadow',
'real entangled shadow' or 'complex GHZ shadow' of a given operator. Using the
notion of probability measure we will denote this restricted by $\mu_{A}^{R}$.




On one hand, for a given matrix $A$ one may study its restricted numerical
shadows $P^R_A(z)$ determined by a given set $R$ of quantum states.
Alternatively, investigating numerical shadows of various matrices of a fixed
dimension $D$ with respect to  a concrete subset $R \in \Omega_D$  one may
analyse its geometry. In this work we study in this way the structure  of the
set of real, separable and maximally entangled quantum states for a two and
three--qubit system.

This paper is organized as follows. In section 2 some basic properties of the
numerical range and the numerical shadow are reviewed. In section 3 we discuss
numerical shadow restricted to real states. In the case of operators acting on
the Hilbert space with a tensor product structure, ${\cal H}={\cal H}_N\otimes
{\cal H}_N$, one defines classes of separable and maximally entangled states.
Numerical ranges restricted to these sets are analysed in section 4 and 5
respectively. In section 6 we show that investigations of trajectories formed by
evolving quantum states projected into the plane of the shadow contributes to
the understanding the dynamics of quantum entanglement. In section 7 we discuss
the simplest case of composite system consisting of three parts and described in
the Hilbert space ${\cal H}_8={\cal H}_2\otimes {\cal H}_2 \otimes {\cal H}_2$.
In the set $\Omega_8$ of pure states acting on this space one defines two
classes of entangled states called $GHZ$ and $W$. For any operator acting on
${\cal H}_8$ it is then natural to introduce the shadow restricted to states
locally equivalent to $GHZ$ or $W$, and these are investigated in section 6.

\section{Standard numerical shadow and the geometry of quantum states} 

\subsection{Classical and quantum states}

Elements of a Hilbert space are used as basic objects of the quantum theory. A
physical system with $D$ distinguishable states can be described by an element
$|\psi\rangle$ of the complex Hilbert space ${\cal H}_D$. It is assumed that
such a \emph{pure quantum state} is normalized, $||\psi||^2=\langle \psi|
\psi\rangle=1$, so it belongs to the hypersphere of dimension $2D-1$. One
identifies any two states, which differ only by a global phase, $|\psi\rangle
\sim |\phi\rangle = e^{\ii \alpha}|\psi\rangle$. The set of all pure quantum
states $\Omega_D$ acting on ${\cal H}_D$ is therefore equivalent to the complex
projective space, $\Omega_D=\CP{D-1}$ (see \emph{e.g.} \cite{BZ06}).

Any convex combination of projectors onto pure states
forms a \emph{mixed quantum state}, 
$\rho=\sum_i p_i |\psi_i\rangle\langle \psi_i|$
with $p_i\ge 0$ and $\sum_i p_i=1$.
Any such state $\rho$, 
also called a density operator, is Hermitian, positive and
normalized by the trace condition, Tr$\rho=1$.
Thus the set ${\cal Q}_D$ of all density matrices 
of order $D$ forms a convex set of $D^2-1$ real dimensions.
The projectors corresponding to the extremal points of the set of density
operators form the set of pure quantum states, $\Omega_D=\CP{D-1} \subset{\cal
Q}_D$. In the one qubit case, $D=2$, the set of pure quantum states forms the
\emph{Bloch sphere}, $\Omega_2=\CP{1}$, which in this case is equivalent to the
boundary of the Bloch ball ${\cal Q}_2$. For a larger dimension the
$(2D-2)$--dimensional set of pure states $\Omega_D$ forms only a zero--measure
subset of the $D^2-2$ dimensional boundary of the set ${\cal Q}_D$.

Note that the definitions of the set of pure and mixed quantum states are
unitarily invariant, so they can be formulated without specifying any basis in
the Hilbert space. On the other hand, among quantum states, one distinguishes
also the set of \emph{classical states}, which are formed by the density
matrices diagonal in a certain basis. Hence any classical state $p$ is
represented by a normalized probability vector, $p=\{ x_1,x_2,\dots, x_D\}$ such
that $x_i\ge 0$ and $\sum_{i=1}^D x_i =1$. The set of classical states forms
thus the probability simplex $\Delta_{D-1} \subset {\mathbbm R}^{D-1}$. Each of
$D$ corners of the simplex represents a classical pure state, and their convex
hull is equivalent to $\Delta_{D-1}$. In the one qubit case, $D=2$, the set
$\Delta_{1}$ of classical states forms an interval which joins two poles of the
Bloch sphere and traverses across the interior of the Bloch ball of one--qubit
mixed quantum states.

\subsection{Numerical range and the set of quantum states}

The definition (\ref{range1}) of the numerical range of a matrix $A$ of order
$D$ specifies $W(A)$ as the set of expectation values $\langle \psi | A| \psi
\rangle$ for normalized pure states $|\psi\rangle$. In fact the compact set
$W(A)$ in the complex plane can be considered as a \emph{projection} of the set
${\cal Q}_D$ of quantum states onto a two plane. The following facts were
recently established in \cite{DGHMPZ11}.

\begin{proposition}\label{prop:classical}
Consider the set of classical states of size $D$, which forms the regular
simplex $\Delta_{D-1}$ in ${\mathbbm R}^{D-1}$.
Then for each normal matrix $A$ of dimension~$D$ there
exists an affine rank $2$ projection $P$ of the set $\Delta_{D-1}$ whose image is
congruent to the numerical range $W(A)$ of the matrix $A$. Conversely for each rank $2$
projection $P$ there exists a normal matrix $A$ whose numerical range $W(A)$
is congruent to image of $\Delta_{D-1}$ under projection $P$.
\end{proposition}

\begin{proposition}\label{prop:quantum}
Let ${\cal Q}_D$ denote the set of quantum states of dimension $D$ embedded in
${\mathbbm R}^{D^2-1}$ with respect to Euclidean geometry induced by
Hilbert-Schmidt distance. Then for each (arbitrary) matrix $A$ of dimension~$D$
there exists an affine rank $2$ projection $P$ of the set ${\cal Q}_D$ whose
image is congruent to the numerical range $W(A)$ of the matrix $A$. Conversely
for each rank $2$ projection $P$ there exists a matrix $A$ whose numerical range
$W(A)$ is congruent to image of ${\cal Q}_D$ under projection $P$.
\end{proposition}

Thus by fixing the dimension $D$ and selecting various operators $A$ of this
order and analysing their numerical ranges $W(A)$ we may gather information
about the projections of the set of quantum states. For instance, for $D=2$ the
projections of the Bloch sphere form ellipses (which could form circles or may
reduce an intervals), and so the numerical range of any operator $A$ of size
$D=2$ may be the result. Similarly the numerical range of operators of size
$D=3$ can be viewed as projections of the set $\Omega_3=\CP{2}$ of single qutrit
quantum states, while the numerical range for matrices of order $D=4$ can be
associated with projections of the set $\Omega_4=\CP{3}$
\cite{DGHPZ11,DGHMPZ11}.

\subsection{Measure on the space of quantum states and the numerical shadow}

The Haar measure on the unitary group $\mathrm{U}(D)$ induces on the set $\Omega_D$ of
quantum pure states the unitarily invariant Fubini-Study measure $\mu$. In the
simplest case of $D=2$ this measure corresponds to the uniform distribution of
points on the Bloch sphere $\Omega_2=\Sphere{2}$.

For any operator $A$ of size $D$ we may compute its expectation value for random
pure state $|\psi\rangle \in \Omega_D$ chosen with respect to measure $\mu$. In
this way for any operator $A$ we define its \emph{numerical shadow} -- a
probability distribution (\ref{shadow}) on the complex plane. Note that by
construction the numerical shadow $P_A(z)$ is supported on the numerical range
$W(A)$. Furthermore, the numerical shadow is unitarily invariant:
$P_A(z)=P_{UAU^{\dagger}}(z)$.

For any normal matrix $A$, which commutes with its adjoint, the numerical shadow
$P_A(z)$ covers the numerical range $W(A)$ with the probability corresponding to
a projection of a regular simplex $\Delta_{D-1}$ of classical states embedded in
${\mathbbm R}^{D-1}$) onto a plane. In general, for a non--normal operator $A$
acting on ${\cal H}_D$, its shadow covers the numerical range $W(A)$ with the
probability corresponding to an orthogonal projection of the complex projective
manifold $\Omega_D=\CP{D-1}$ onto a plane.

In this work we will analyse the numerical range restricted \cite{GPMSCZ09} to a
certain subsets $R$ of pure quantum states and the corresponding
\emph{restricted} numerical shadows, with probability density determined by
random states distributed on the subset $R$ according to the measure induced by
the Fubini--Study measure on $\Omega_D$.

\subsection{Standard numerical shadow for a diagonal matrix}

Consider a diagonal matrix of size $D$, namely   $X={\rm diag}(x_1, \dots x_D)$.
Let $|1\rangle$ be an arbitrary fixed pure state in ${\cal H}_D$.
Then the entire set of random pure states can be obtained as
$|\psi\rangle = U|1\rangle$, where $U\in \mathrm{U}(D)$ is a random unitary 
matrix distributed according the Haar measure. Thus the expansion coefficients
of the state $|\psi\rangle$ read $(U_{11}, U_{12},\dots, U_{1D})$. 

The \emph{numerical shadow} of the operator $X$ is defined as the density
distribution of random numbers $z:=\langle \psi|X|\psi\rangle$, where
$|\psi\rangle$ is a random state defined by the random unitary matrix $U$.
Therefore
\begin{equation}
z:=\langle 1|U^{\dagger} X U |1\rangle = \sum_{i=1}^D x_i |U_{1i}|^2 =x\cdot q
\end{equation}

Note that the vector of coefficients $q={q_1,\dots q_D}$ with $q_i=|U_{ij}|^2$
belongs to the $D-1$ dimensional simplex $\Delta_{D-1}$. So the complex random
variable $z$ can be treated as a scalar product, $z=x\cdot q$ where the complex
vector $x$ is given by the diagonal of the operator $X$, while the probability
vector $q$ is random. Its distribution inside the simplex depends on the
distribution in the space of unitary matrices $\mathrm{U}(D)$. In cases of interest for
us this distribution belongs to the class of \emph{Dirichlet distributions}
parametrized by a real number $s$, 
\begin{equation}
P_s(q) \ \propto  \ \delta(\sum_{i=1}^D q_i -1)\ 
               (q_1 q_2 \cdots q_D)^{s-1} \ .
\label{dirichlet}
\end{equation}

In particular, we are interested in two situations.
\begin{itemize}
\item[A)] \emph{Standard (complex) numerical shadow} \cite{DGHPZ11,DGHMPZ11},
     generated by the Haar measure on $\mathrm{U}(D)$.
   Then the vector $q$ is distributed uniformly with respect to the flat
   (Lebesgue) measure on $\Delta_{D-1}$, 
  i.e.  the Dirichlet measure with the Dirichlet parameter $s=1$ \cite{BZ06}.

\item[B)]
\emph{Real} numerical shadow, generated by the Haar measure on the group  
  $\mathrm{O}(N)$ of orthogonal matrices.
   Then the vector $p$ is distributed with respect to the \emph{statistical measure},
   $P_{1/2}(q) \propto 1/\sqrt{q_1 \cdots q_D}$,  
  i.e.  the Dirichlet measure with the Dirichlet parameter $s=1/2$. 
\end{itemize}

In both cases the shadow covers the entire numerical range of $X$, equal to the
convex hull of the spectrum $\{x_i\}_{i=1}^D$ with a density determined by the
appropriate Dirichlet distribution. The case of real numerical shadow, i.e. the
shadow with respect to the real pure states is treated in more detail in the
subsequent section.

\section{Numerical shadow with respect to real states}

Quantum states belonging to a complex Hilbert space form a standard tool of the
quantum theory. However, in some cases it is instructive to restrict attention
to real states only. On one hand the set of the real states is easier to analyse
than the full set of complex states of larger dimensionality. For instance, the
phenomenon of quantum entanglement can be studied for the case of real states of
a four--level systems, sometimes referred to as a pair of rebits (i.e. real
bits) \cite{CFR01}.

On the other hand is some physical applications it is easier to use real
orthogonal rotation matrices to construct elements of the entire set of real
density matrices. Therefore in this section we will study the shadow of an
operator with respect to the set of quantum states with real coefficients. Such
an investigation allows us to improve understanding of the structure of the set
of real mixed states and its natural subsets.

\subsection{Real shadow of operators of size $D=2,3,4$}

Consider the subset of pure quantum states of size $D$, that can be represented
by real expansion coefficients in a given basis. This set forms a $(D-1)$
dimensional real projective space $\RP{D-1} \subset \CP{D-1}$.

Substituting into the definition (\ref{rest_range}) of a restricted numerical
range \cite{SHDHG08,GPMSCZ09} the set $\RP{D-1} \subset \CP{D-1}$ as the subset
$R$ of $\Omega_D$ we arrive at the numerical range restricted to real states. In
general the restricted numerical range needs not to be convex or simply
connected. Similarly, restricting the integration in (\ref{rshadow}) to the set
$\RP{D-1}$ of real quantum states 
we obtain the shadow of the operator $A$ with respect to real states. For
brevity we will also use a shorter expression, the \emph{real shadow} of $A$.

In the case of $D=2$ the set of real states forms the real projective space
equivalent to a circle, $\RP{1} \sim \Sphere{1}$. Thus the real shadow of a
generic, non-normal matrix of size two forms a singular probability distribution
supported on an elliptic curve in the complex plane, and not inside its
interior.

\begin{figure}[ht!]
\begin{center}
\newlength{\wdth}
\setlength{\wdth}{0.3\textwidth}
\subfigure[\ $A_{2}$]{
\includegraphics[width=\wdth]{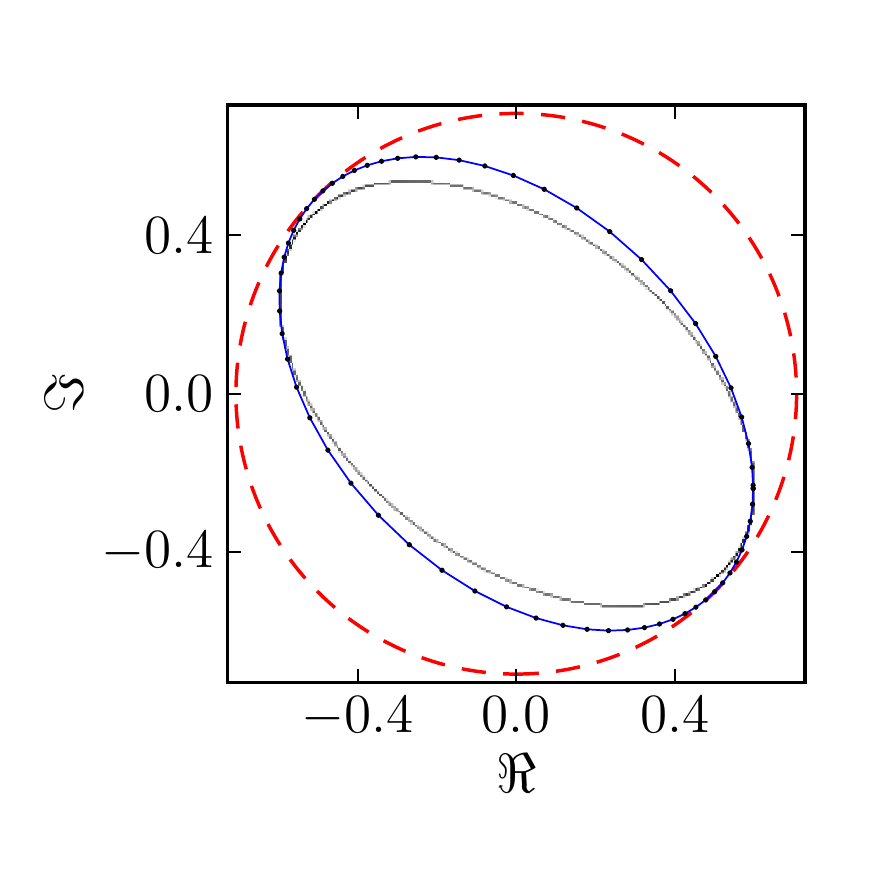}
\label{fig:real-S3A2}
}
\subfigure[\ $A_{3a}$]{
\includegraphics[width=\wdth]{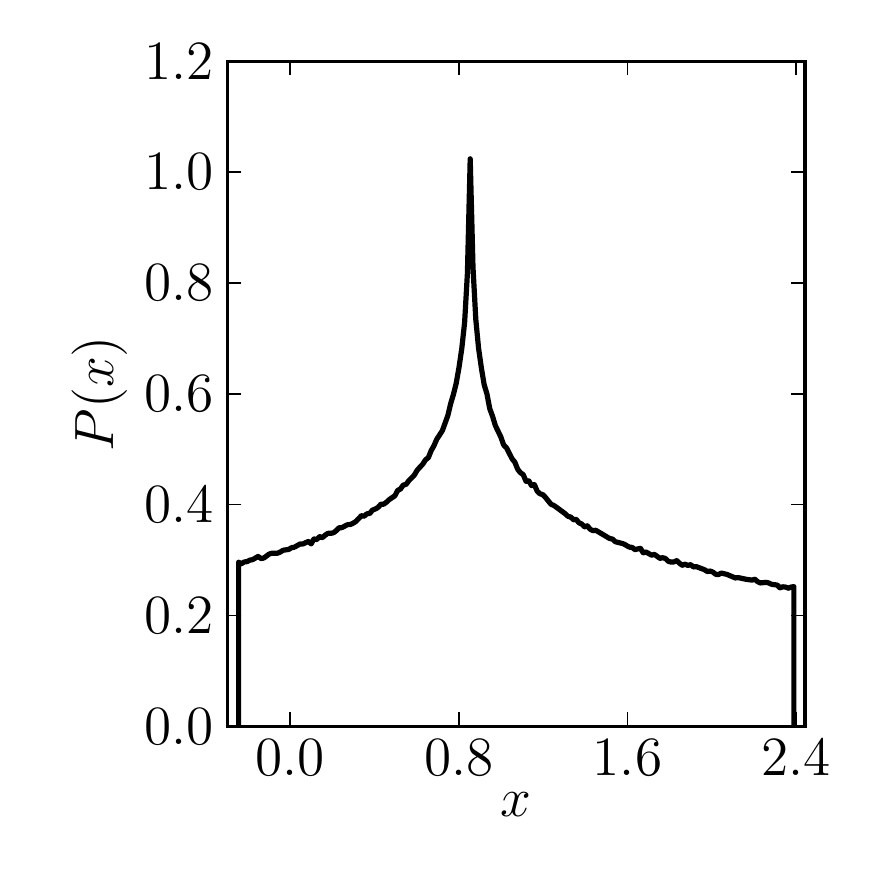}
\label{fig:real-S3A3a}
}
\subfigure[\ $A_{3b}$]{
\includegraphics[width=\wdth]{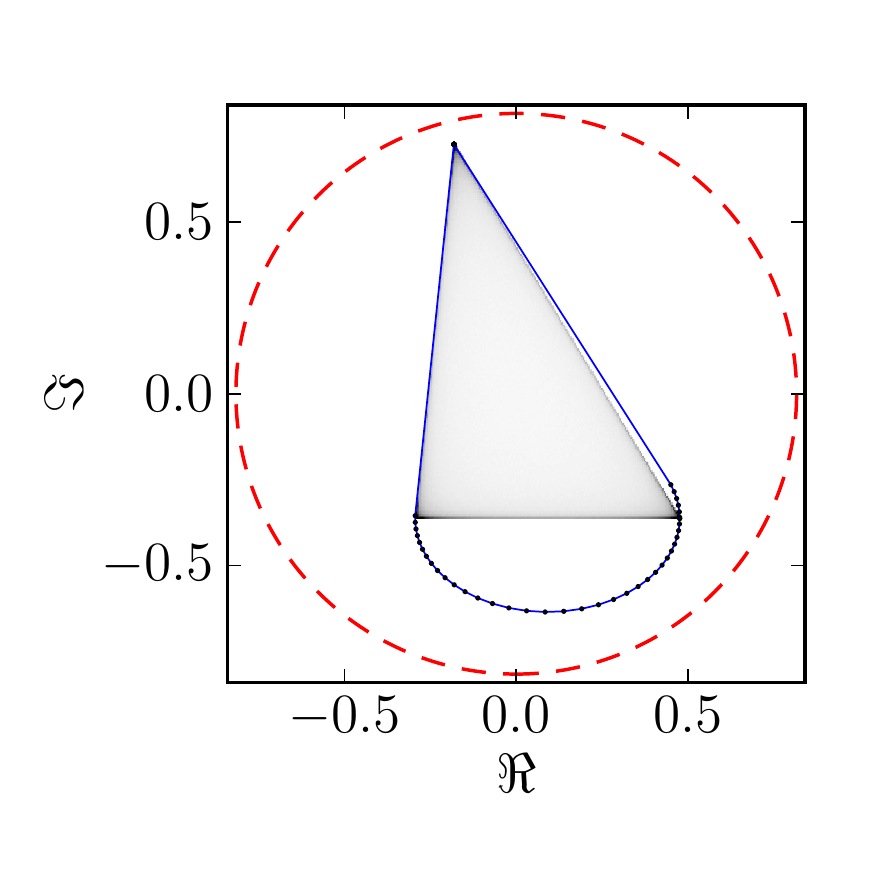}
\label{fig:real-S3A3b}
}\\
\subfigure[\ $A_{4a}$]{
\includegraphics[width=\wdth]{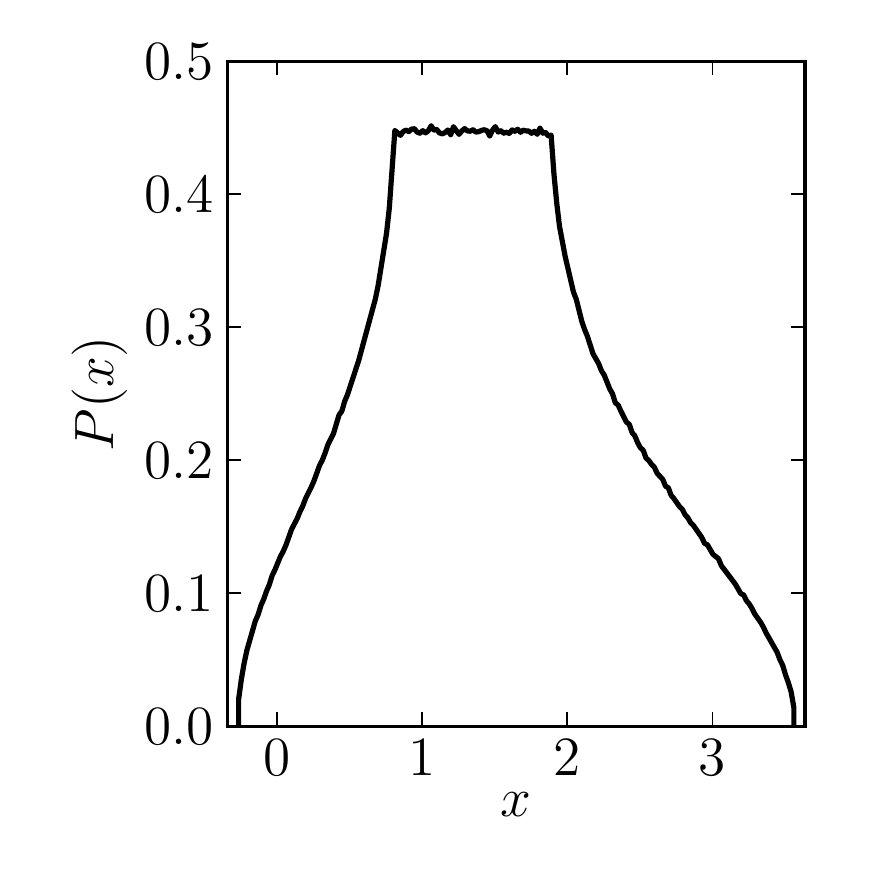}
\label{fig:real-S3A4a}
}
\subfigure[\ $A_{4b}$]{
\includegraphics[width=\wdth]{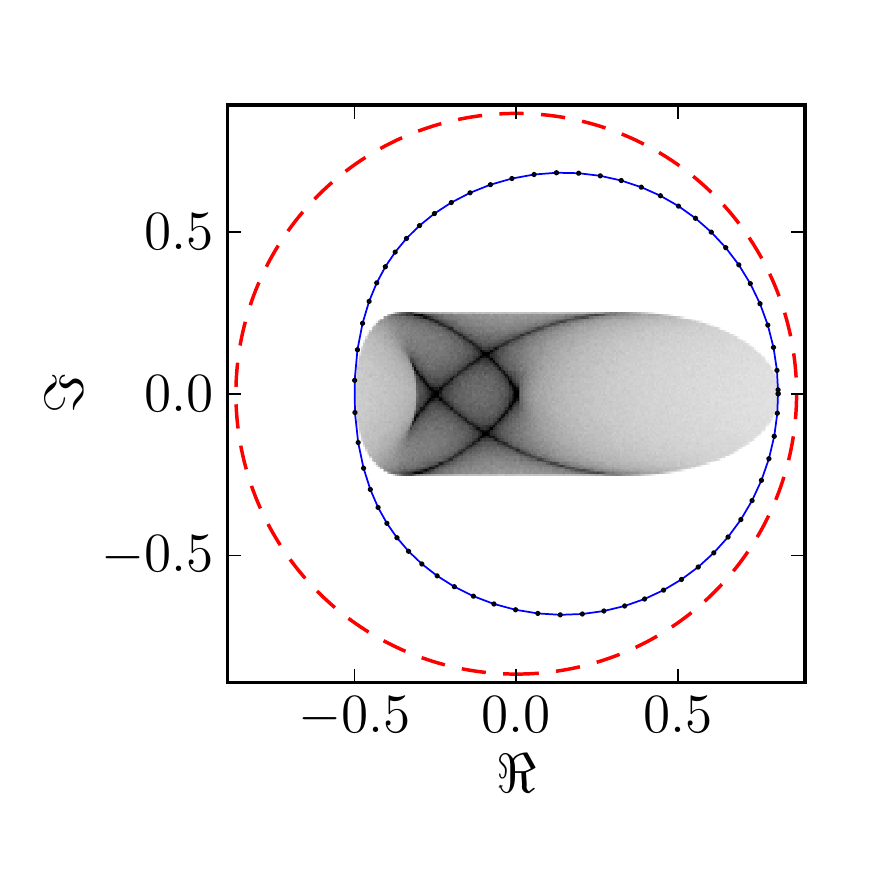}
\label{fig:real-S3A4b}
}
\caption{Numerical shadow with respect to real states
of illustrative operator of size $D=2$ --- panel \subref{fig:real-S3A2}, 
represents projection onto the plane of real projective space $\RP{2}$,
$D=3$ --- panels \subref{fig:real-S3A3a} and \subref{fig:real-S3A3b} represent 
projections onto 
the plane of real 
projective space $\RP{3}$, 
while the real shadows of operators of $D=4$, 
panels \subref{fig:real-S3A4a} and \subref{fig:real-S3A4b} 
represent projections of real projective space $\RP{4}$.
Dashed circles represent the image of the outsphere 
in which the set of quantum states is inscribed
while dotted lines denote the standard numerical range.
Plots \subref{fig:real-S3A2}, \subref{fig:real-S3A3b} and \subref{fig:real-S3A4b} are done for matrices translated in such a way
that their traces are equal to zero and suitably rescaled as described in \cite{DGHMPZ11}.
}
\label{fig:shadreal}
\end{center}
\end{figure}

The real shadows presented in Fig. \ref{fig:shadreal}
are obtained for illustrative operators of size $D=2$
\begin{equation}
A_{2} = \left[\begin{array}{cc}
-1  &  1-\ii \\
-\ii  &  1 
\end{array}\right] ,
\label{a2}
\end{equation}
$D=3$, 
\begin{equation}
A_{3a} = \left[\begin{array}{ccc}
0  &  1 & 1 \\
0  &  1 & 1  \\
0  &  0 & 2 
\end{array}\right] ,  \quad
A_{3b} = \left[\begin{array}{ccc}
0  &  1 & 0  \\
0  &  1 & 0  \\
0  &  0 & 2\ii
\end{array}\right] ,
\label{a3}
\end{equation}
and $D=4$ 
\begin{equation}
A_{4a} = \left[\begin{array}{cccc}
0  &  1 & 1 & 1 \\
0  &  1 & 1 & 1 \\
0  &  0 & 2 & 1 \\
0  &  0 & 0 & 3 
\end{array}\right] ,  \quad
A_{4b} = \left[\begin{array}{cccc}
0  &  1  & 1 & 1 \\
0  &  1  & 1 & 1 \\
0  &  0  & \ii & 1 \\
0  &  0  & 0 & 1+\ii 
\end{array}\right] .
\label{a4}
\end{equation}

These probability distributions can be thus interpreted as shadows of real
projective spaces $\RP{2}$, $\RP{3}$ and $\RP{4}$ on a plane. In the case \ref{fig:real-S3A3a} and
\ref{fig:real-S3A4a} the shadow is supported on a real line, so we plot the corresponding
probability distribution $P_A^{\mathbbm R}(x)$. In the other cases \ref{fig:real-S3A2}, \ref{fig:real-S3A3b} and \ref{fig:real-S3A4b}
the real shadow is supported on the complex plane, so the density is encoded in
the grey scale. The circle drawn by red dotted line represents the image of the
sphere of dimension $D^2-1$, in which the set of quantum states can be
inscribed. The blue dotted line represents the boundary of the standard
numerical range of an operator. The restricted shadow is supported in this set
or its subset.

\begin{figure}[ht!]
\begin{center}
    \setlength{\wdth}{0.3\textwidth}
    \subfigure[\ $\mathrm{diag}\left(\left\{e^{2 \pi \ii k /4}\right\}_{k=1}^{4} \right)$]{
    \includegraphics[width=\wdth]{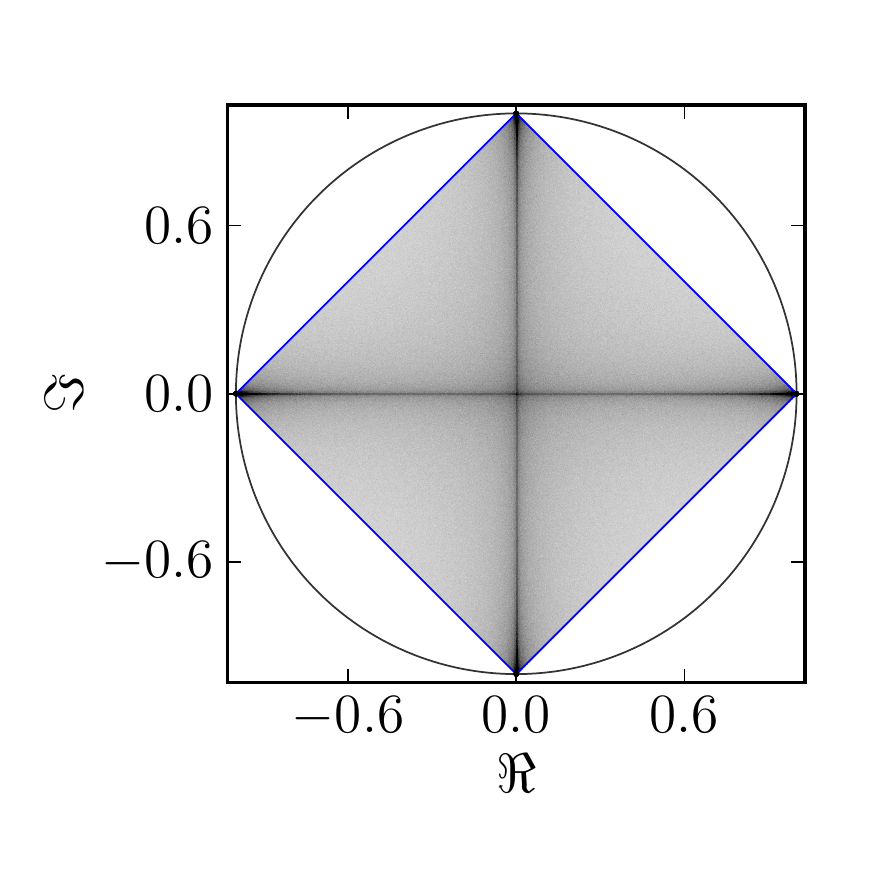}
    \label{fig:real-S3D4}
    }
    \subfigure[\ $\mathrm{diag}\left(\left\{e^{2 \pi \ii k /5}\right\}_{k=1}^{5} \right)$]{
    \includegraphics[width=\wdth]{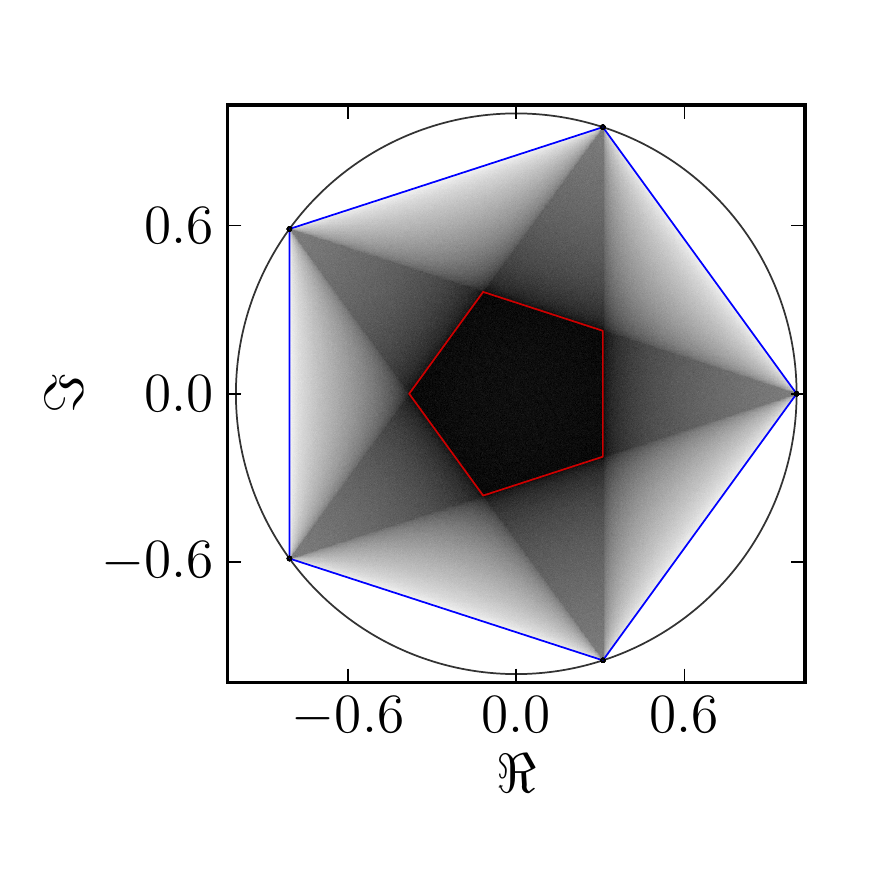}
    \label{fig:real-S3D5}
    }
    \subfigure[\ $\mathrm{diag}\left(\left\{e^{2 \pi \ii k /7}\right\}_{k=1}^{7} \right)$]{
    \includegraphics[width=\wdth]{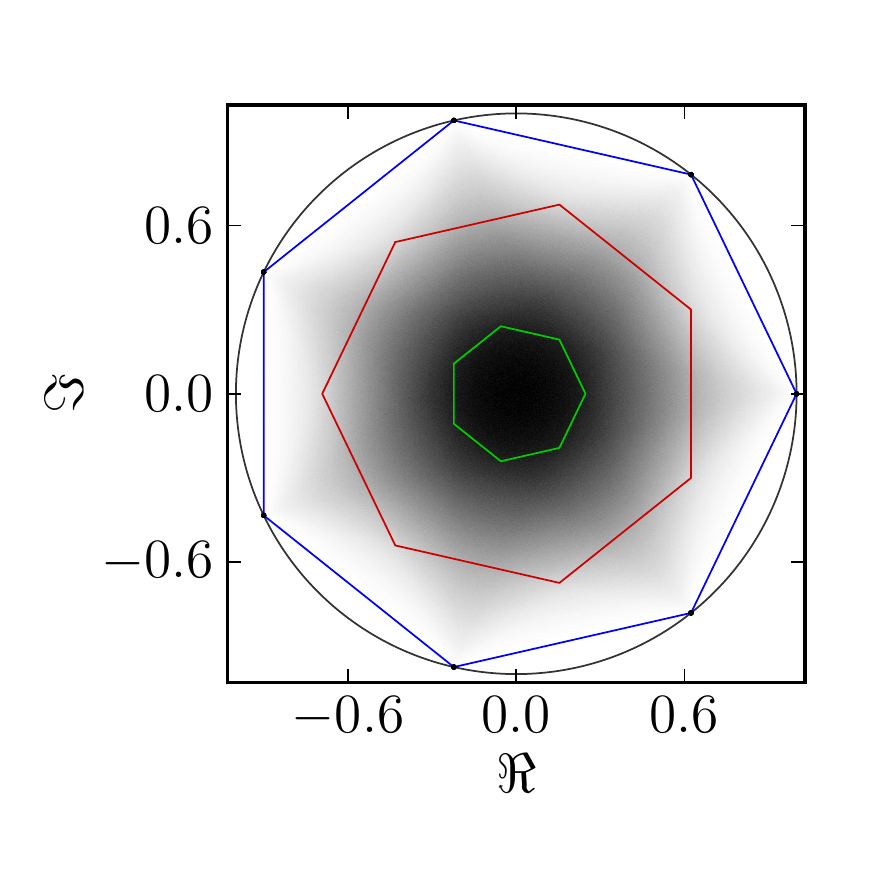}
    \label{fig:real-S3D7}
    }
    \caption{Real shadows of diagonal unitary matrices $U$ of size $D$ having
    eigenvalues evenly distributed on the unit circle, \subref{fig:real-S3D4}
    $D=4$, \subref{fig:real-S3D5} $D=5$ and \subref{fig:real-S3D7} $D=7$. The
    shadow is supported in the regular polygon -- the numerical range
    $\Lambda_1(U)=W(U)$. Inner polygons represent higher order numerical ranges
    $\Lambda_2(U)$ and $\Lambda_3(U)$.}
    \label{fig:stars}
\end{center}
\end{figure}

The numerical shadow carries also some information about the higher rank
numerical range \cite{CKZ06,CHKZ07} of an operator. For instance, the darker
area of the shadow corresponding to a larger probability \cite{DGHMPZ11} allows
one to recognize in Fig.~\ref{fig:stars} the \emph{numerical range of rank $2$},
written $\Lambda_2(U)$ of selected unitary matrices. In the case of $D=4$ shown
in Fig.~\ref{fig:stars} \subref{fig:real-S3D4} it is equal to a single point at
which the two diagonals of the quadrangle cross, while for the case $D=5$ shown
in panel \subref{fig:real-S3D5} the numerical range of rank two is represented
by the inner pentagon located inside the numerical range in this case numerical
shadow appears to be uniform in the set $\Lambda_2(U_5)$. In the case $D=7$
shown in the panel \subref{fig:real-S3D7} one can see darker areas between
segments connecting every other eigenvalue.

\subsection{A general approach to real shadow}

Given $M\in M_N(\mathbb{C})$, the real shadow is taken to mean the distribution
of $(Mx,x)$ when $x\in\mathbb{R}^N$ is a unit vector randomly chosen from the
uniform distribution on $S^{N-1}=\{x\in\mathbb{R}^N:\|x\|=1\}$.

If $M$ itself is real, we have $(Mx,x)=(x,Mx)$ so that the distribution of
$(Mx,x)$ is the same as that of the symmetric $(M+M^T)/2$. Since the uniform
distribution on $S^{N-1}$ is invariant under orthogonal transformations, we are
free to diagonalize this symmetric matrix. In other words, the real shadow of
$M\in M_N(\mathbb{R})$ is that of $\mathrm{diag}(a_1,\dots,a_N)$, where $a_k$
are the (real) eigenvalues of $(M+M^T)/2$. Thus we need the distribution of
\begin{equation}\label{eqn:real-john-e1}
\sum_{k=1}^Na_kx_k^2.
\end{equation}

It is known that $x$ uniform on $S^{N-1}$ implies that $p=(x_1^2,\dots,x_n^2)$
has a Dirichlet--(1/2) distribution on the simplex
$\Delta_{N-1}=\{p\in[0,1]^N:\sum_1^N p_k=1\}$, \emph{ie.} its density is
proportional to $1/\sqrt{p_1p_2\dots p_N}$. 

The distribution of (\ref{eqn:real-john-e1}) is known in general \cite{PC00},
so for $a_N<a_{N-1}<\dots<a_1$ we have
\begin{equation}
\label{eqn:real-john-e3}
\mathrm{Prob}\left(\sum_{k=1}^Na_kx_k^2\leq t\right)=\frac12-\frac1\pi\int_0^\infty
\frac{\sin \bigl[ \frac12\sum_{j=1}^N\tan^{-1}\bigl( (a_j-t)u \bigr) \bigr]}
{u\prod_{j=1}^N[1+(a_j-t)^2u^2]^\frac14}\,du.
\end{equation}

As an example, let us consider the Jordan nilpotent 
$
J_3 = \left(
\begin{smallmatrix}
0 & 1 & 0\\ 
0 & 0 & 1\\ 
0 & 0 & 0
\end{smallmatrix}
\right).
$
For the real shadow
of $J_3$, we first note that it is the same as that of
$\mathrm{diag}(1/\sqrt2,0,-1/\sqrt2)$. Rescaling for convenience, we compute the
real shadow of $\mathrm{diag}(1,0,-1)$. The density $g_3(t)$ of $p_1-p_3$ is 0
outside $[-1,1]$, satisfies $g_3(-t)=g_3(t)$, and, in view of the
Dirichlet--(1/2) distribution on $\Delta_2$, $g_3(t)$ is proportional to
\begin{equation}\label{eqn:real-john-e2}
\int_0^{\frac{1-t}2}\frac{dx}{\sqrt{(x+t)(1-t-2x)x}},
\end{equation}
for $0<t<1$. In particular, $g_3(0)=+\infty$. 
In \cite{PC00} the case $N=3$ is treated in terms of hypergeometric
functions and we can compute
\begin{equation}\label{eqn:real-john-e4}
g_3(t)=\frac1{2\sqrt{2t}}\,\, 
{}_2F_1\left(\frac12,\frac12;1,\frac{t-1}{2t}\right),
\end{equation}
for $t\in(0,1)$.

For a general $M\in M_N(\mathbb{C})$, one may write $M=A+\ii B$ with $A,B\in
M_N(\mathbb{R})$, and by diagonalizing $(A+A^T)/2$ and $(B+B^T)/2$ we may
understand the distributions of $(Ax,x)$ and $(Bx,x)$ separately, using the
techniques above. Usually the \emph{joint} distribution will be obscure, but we
may have some hope of understanding it in certain cases. If, for example, $M$ 
is normal and presented in diagonal form, it may be possible to see how the
horizontal and vertical distributions knit together. 

\begin{figure}[ht]
	\setlength{\wdth}{0.3\textwidth}
	\centering
	\subfigure[\ Complex shadow (left plot) and real shadow (right plot) of a Hermitian matrix $\mathrm{diag}(0,1,3,5)$.]{
	\includegraphics[width=\wdth]{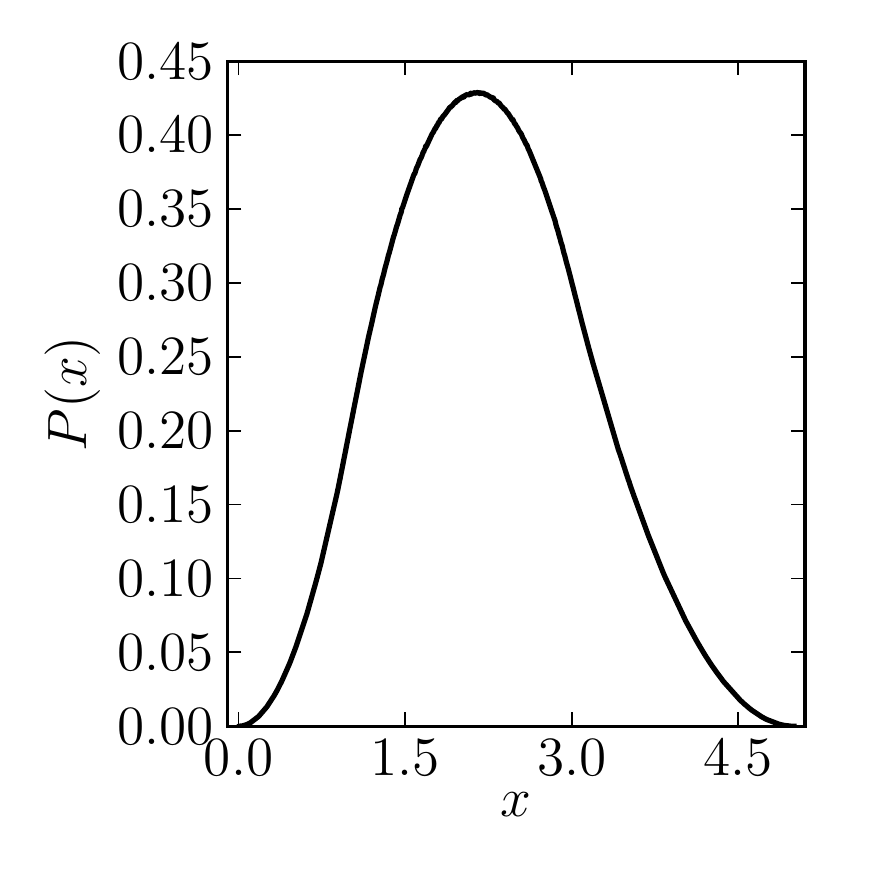}
	\includegraphics[width=\wdth]{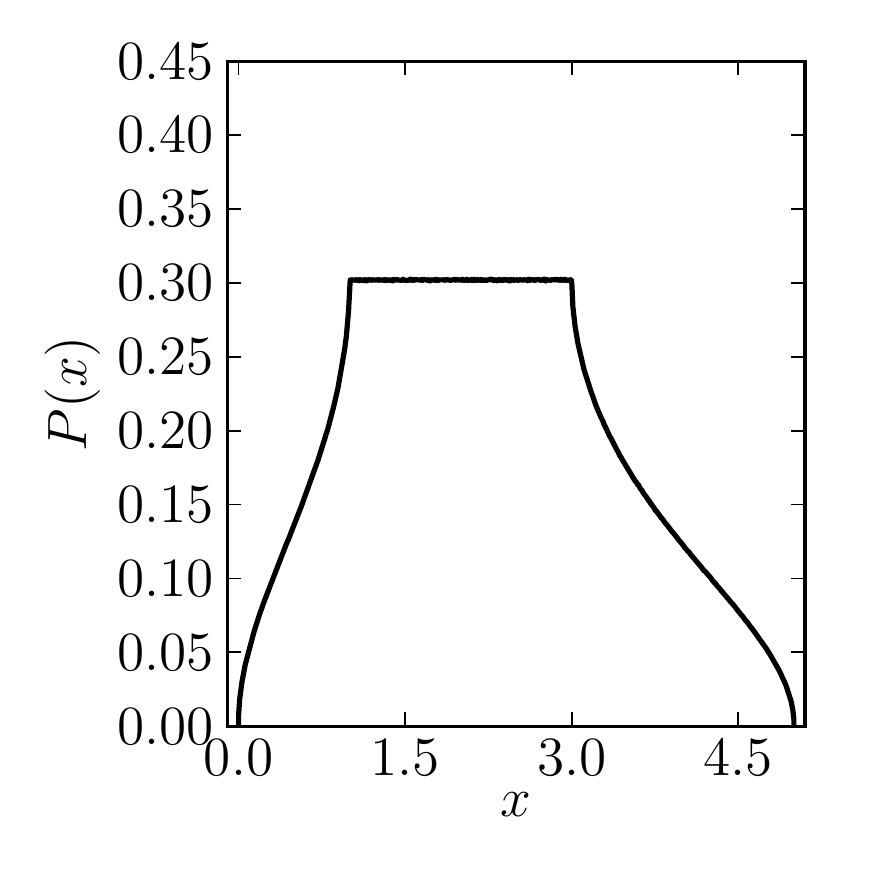}
	\label{fig:complex_and_real_D3}
	}	
	\\
	\subfigure[\ Complex shadow (left plot) and real shadow (right plot) of a unitary matrix $\mathrm{diag}(e^{\ii \frac{\pi}{8}},
			e^{\ii \frac{3\pi}{8}},
			e^{\ii \frac{7\pi}{8}},
			e^{\ii \frac{11\pi}{8}})$ .
	]{
	\includegraphics[width=\wdth]{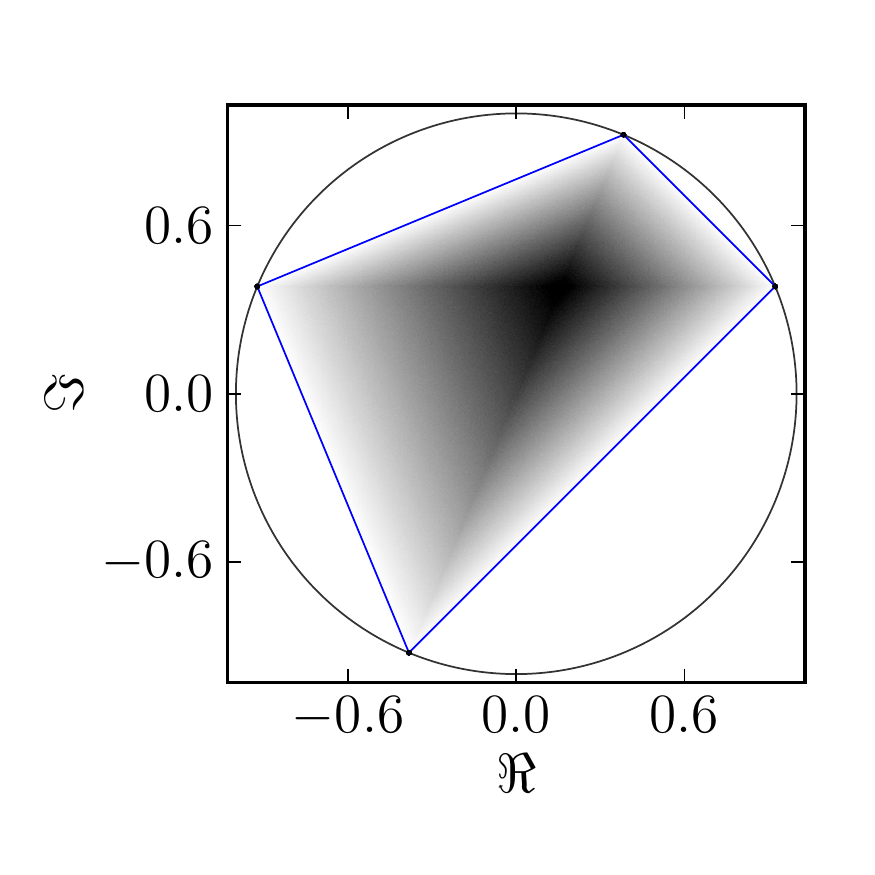}
	\includegraphics[width=\wdth]{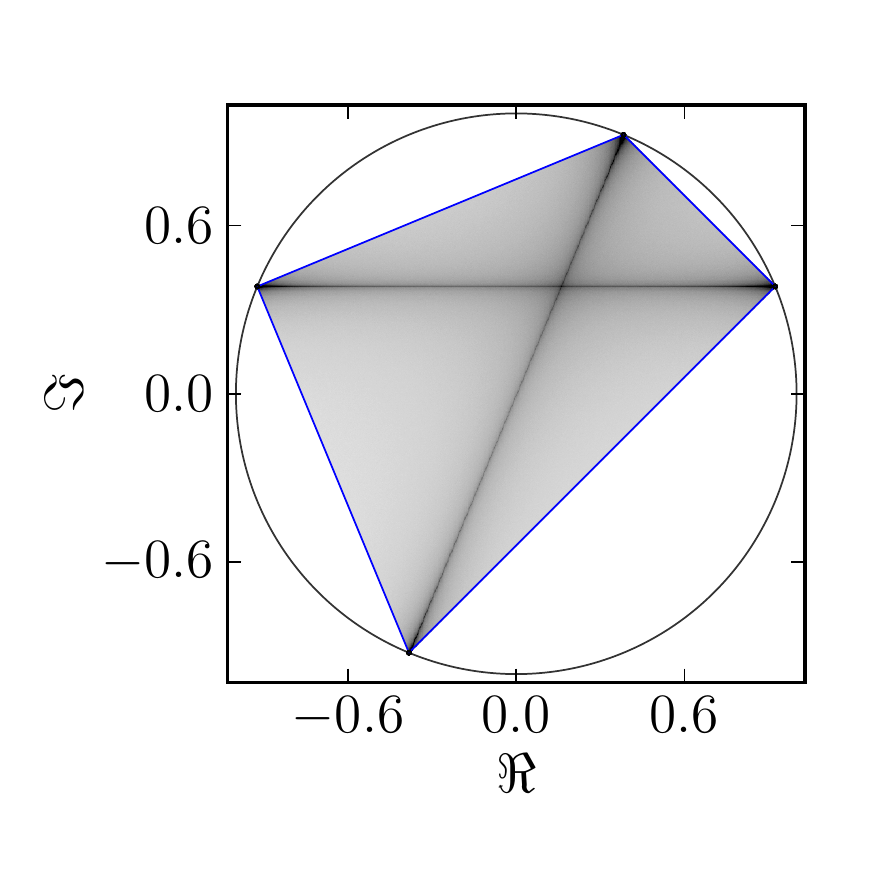}
	\label{fig:complex_and_real_Vphi3}
	}
	\caption{Comparison of real and complex numerical shadow for Hermitian
	(panel \ref{fig:complex_and_real_D3}) and unitary (panel
	\ref{fig:complex_and_real_Vphi3}) matrices.
	}
	\label{fig:john-example}
\end{figure}

\begin{figure}
	\setlength{\wdth}{0.3\textwidth}
	\centering
	\subfigure{\includegraphics[width=\wdth]{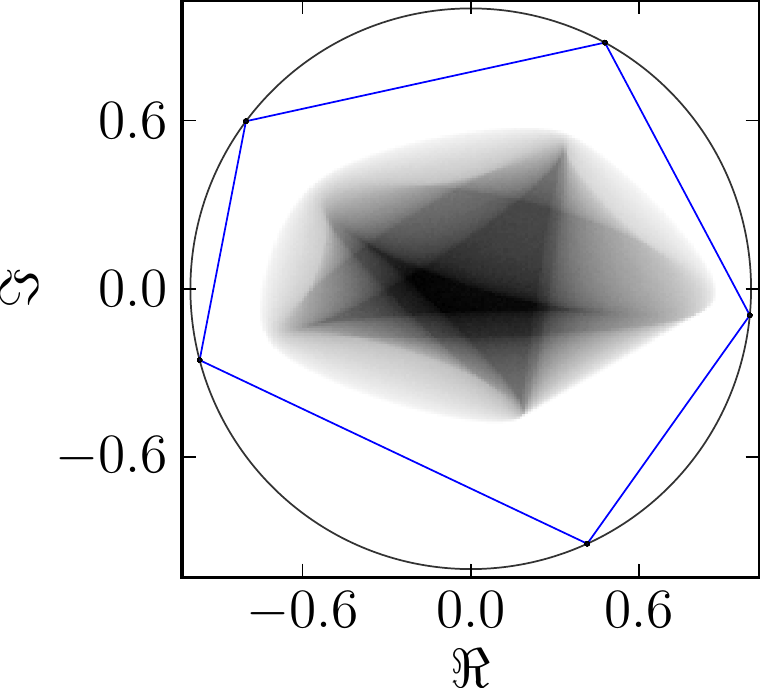}
	\label{fig:random_state_real_U57}
	}
	\subfigure{\includegraphics[width=\wdth]{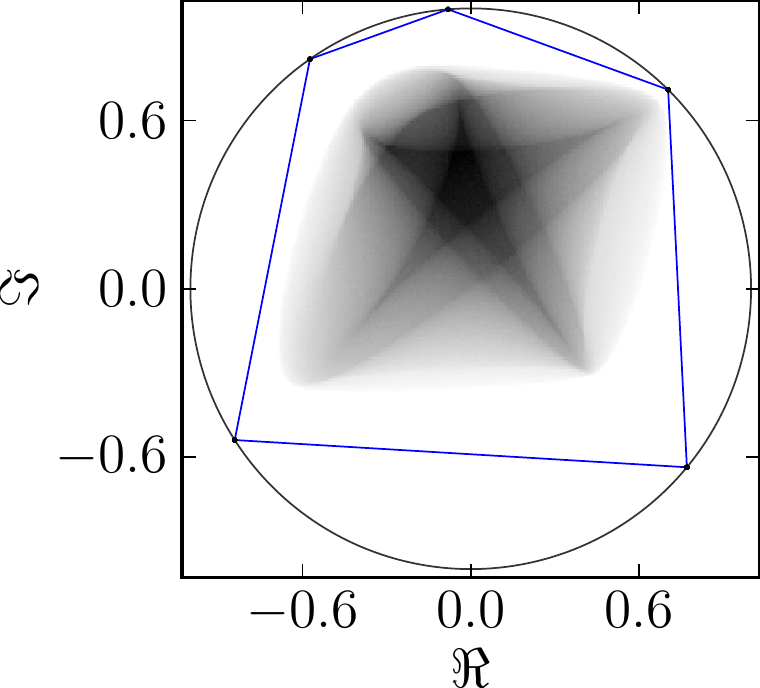}
	\label{fig:random_state_real_U517}
	}
	\caption{Real shadow of two generic unitary matrices of dimension five is
	supported on the subset of the numerical range formed by the pentagon
	plotted.}
	\label{fig:unitary5real}
\end{figure}

Fig.~\ref{fig:john-example} shows various shadows of $M$ when $M$ is a
$4\times4$ unitary (in diagonal form). The sort of 1D shadow density seen in the left panel of
Fig.~\ref{fig:complex_and_real_D3} seems typical; notably, the real density of a $4\times4$
Hermitian is constant between the two middle eigenvalues. Presumably, the
vertical density has a similar form. Fig.~\ref{fig:unitary5real} presents two
examples of real shadow of generic unitary matrices of dimension five. Support
of the shadow does not coincide with numerical range of those matrices because
a~generic unitary matrix can not be diagonalized using only orthogonal matrices.

\subsection{Real shadows and their moments} 
Since methods based on moments of the shadow distribution were so effective in
the complex case,  we may try to mimic them for the real shadow.  It is not hard
to adapt the method outlined at the beginning of the proof of Proposition 5.1
from \cite{DGHPZ11} to show that, for $\alpha\in\mathbb{N}_0^N$
\[
R(\alpha)=\int_{S^{N-1}} x^\alpha = 0
\]
unless each component of $\alpha$ is even, and that
\begin{equation}\label{eqn:real-john-e5}
R(2\beta)=\frac{\prod_{k=1}^N(\frac12)_{\beta_k}}{(\frac{N}2)_{|\beta|}}.
\end{equation}
Here we use the shifted factorial notation: $(x)_n=x(x+1)\dots(x+n-1)$, with the
convention $(x)_0=1$; also $|\beta|$ denotes $\sum_1^N\beta_k$.

In principle, we can use Eq.~(\ref{eqn:real-john-e5}) to evaluate the moments of
the real shadow density $g(t)$ of $\mathrm{diag}(a_1,\dots,a_N)$
\begin{equation}\label{eqn:real-john-e6}
\int t^ng(t)\,dt=\int_{S^{N-1}} \left(\sum_k a_kx_k^2\right)^n = 
\sum_{|\beta|=n} \binom{n}{\beta}a^\beta R(2\beta).
\end{equation}
In the complex case we were able to progress beyond this point, obtaining such
effective relations for the moments as the \emph{determinant relation} (see:
Eq.~(10) in \cite{DGHPZ11}).

As a test case, we may try to use Eq.~(\ref{eqn:real-john-e6}) to find the
moments of $g_3(t)$, the real shadow density of $\mathrm{diag}(1,0,-1)$,
discussed in the previous example. We have
\[
\int_{-1}^1 t^ng_3(t)\,dt=\int_{S^2}(x_1^2-x_3^2)^n = \sum_{k=0}^n \binom{n}{k}(-1)^k R(2(n-k),0,2k)
\]
and finally we obtain that
\begin{equation}\label{eqn:real-john-e7}
\int_{-1}^1 t^ng_3(t)\,dt=\sum_0^n\binom{n}{k}(-1)^k\frac{(\frac12)_{n-k}(\frac12)_k}{(\frac32)_n}.
\end{equation}
It can be shown, that for even $n=2m$ we have
\[
\int_{-1}^1 t^ng_3(t)\,dt = \frac{(2m+2)(2m+4)\dots(4m)}{(2m+1)(2m+3)\dots(4m+1)}.
\]
To see this we start with equation Eq.~(\ref{eqn:real-john-e7}).
Changing the index of summation from $k$ to $n-k$ shows that the sum equals
$\left(-1\right)^{n}$ times itself, hence equals zero for odd $n$. Now
suppose that $n=2m,m=0,1,2,\ldots$. The sum can be written in hypergeometric
form%
\[
\int_{-1}^{1}t^{n}g_{3}\left(  t\right)  dt = 
\frac{\left(\frac{1}{2}\right)_{2m}}{\left(\frac{3}{2}\right)_{2m}}
\sum_{k=0}^{2m}
 \frac{\left(-2m\right)_{k}\left(\frac{1}{2}\right)_{k}}
      {k!\left(\frac{1}{2}-2m\right)_{k}}
 \left(-1\right)^{k} =
 \frac{\left(\frac{1}{2}\right)_{2m}}{\left(\frac{3}{2}\right)_{2m}}
 {}_{2}F_{1} \left(-2m,\frac{1}{2};1-2m-\frac{1}{2};-1\right).
\]
Kummer's summation formula (see \cite[p. 10]{Bailey}) implies 
(later we set $a=-2m,b=\frac{1}{2}$)
\[
_{2}F_{1}\left(  a,b;1+a-b;-1\right)  =2^{-a}~_{2}F_{1}\left(  \frac{a}%
{2},\frac{a+1}{2}-b;1+a-b;1\right).
\]
Now we set $a = -2m $ and use the terminating form of Gauss's sum (the
Chu-Vandermonde formula)
to obtain%
\begin{align*}
_{2}F_{1}\left(  -2m,b;1-2m-b;-1\right)   &  =2^{2m}~_{2}F_{1}\left(
-m,-m+\frac{1}{2}-b;1-2m-b;1\right) \\
&  =2^{2m}\frac{\left(  \frac{1}{2}-m\right)  _{m}}{\left(  1-2m-b\right)
_{m}}=2^{2m}\frac{\left(  \frac{1}{2}\right)  _{m}}{\left(  b+m\right)  _{m}}.
\end{align*}
Thus, setting $b = \frac{1}{2}$ we obtain, that the integral equals%
\[
\int_{-1}^{1}t^{n}g_{3}\left(  t\right)  dt= 
\frac{\frac{1}{2}}{\frac{1}{2}+2m}2^{2m}\frac{\left(  \frac{1}{2}\right)
_{m}}{\left(  \frac{1}{2}+m\right)  _{m}}=2^{m}\frac{\left(  2m\right)
!}{m!\left(  2m+1\right)  \left(  2m+3\right)  \ldots\left(  4m+1\right)  }.
\]



We use the variance of a complex random variable to give quantitative insight
into Figs.~\ref{fig:stars}-\ref{fig:unitary5real}. The calculations leading to the following formulae are based
on the values of integrals of monomials over the real or complex unit spheres.
Suppose $A$ is a normal matrix, written in the form $A=VDV^{\dag}$ where $V$
is unitary and $D$ is diagonal with entries $\lambda_{1},\lambda_{2}%
,\ldots,\lambda_{N}$. Thus the columns of $V$ are eigenvectors of $A$. Let $X$
denote the random variable $\left\langle \psi|A|\psi\right\rangle $ and let
$m=\frac{1}{N}trA=\frac{1}{N}\sum_{j=1}^{N}\lambda_{j}$. If $\psi$ is a random
vector in the unit sphere in $\mathbb{R}^{N}$ or $\mathbb{C}^{N}$ then
$\mathcal{E}X=m$ (where $\mathcal{E}$ denotes the expectation). Denote the
\textit{complex variance }$v\left(  X\right)  =\mathcal{E}\left(  \left\vert
X-m\right\vert ^{2}\right)  $, a measure of (2-dimensional) spread of $X$. For
the complex shadow one finds
\[
v\left(  X\right)  =\frac{1}{N\left(  N+1\right)  }\sum_{j=1}^{N}\left\vert
\lambda_{j}-m\right\vert ^{2}.
\]
For the real shadow and when $A$ can be diagonalized by a real orthogonal
matrix, that is, $V$ is orthogonal ($VV^{T}=I$) then
\[
v\left(  X\right)  =\frac{2}{N\left(  N+2\right)  }\sum_{j=1}^{N}\left\vert
\lambda_{j}-m\right\vert ^{2},
\]
which is larger than the variance of the complex shadow (see Fig.
\ref{fig:complex_and_real_Vphi3} ). In the general real case $V$ is not
orthogonal, then set
$C\left(  V\right)  _{ij}=\left\vert \sum_{k=1}^{N}V_{ki}V_{kj}\right\vert
^{2}$ for $1\leq i,j\leq N$. Thus $C\left(  V\right)  $ is a symmetric
unistochastic matrix, and its eigenvalues all lie in $\left[  -1,1\right]  $.
By straightforward computations we find%
\[
v\left(  X\right)  =\frac{1}{N\left(  N+2\right)  }\sum_{i,j=1}^{N}\left(
\lambda_{i}-m\right)  \overline{\left(  \lambda_{j}-m\right)  }\left(
\delta_{ij}+C\left(  V\right)  _{ij}\right)  ,
\]
a positive quadratic form in $\left(  \lambda_{i}-m\right)  _{i=1}^{N}$. The
eigenvalues of $I+C\left(  V\right)  $ lie in $\left[  0,2\right]  $, and all
equal $2$ when $V$ is real orthogonal.

As example consider the unitary Fourier matrix whose entries are primitive $N^{th}$
roots of unity : $V_{jk}=\frac{1}{\sqrt{N}}\exp\left(  \frac{2\pi i}{N}\left(
j-1\right)  \left(  k-1\right)  \right)  ,1\leq j,k\leq N$. Then $C\left(
V\right)  _{jk}=1$ if $j+k\equiv2\operatorname{mod}N$ otherwise $C\left(
V\right)  _{jk}=0$. The eigenvalues of $I+C\left(  V\right)  $ are $2$ with
multiplicity $\left\lfloor  \frac{N}{2}\right\rfloor  +1$ and $0$ with multiplicity
$\left\lfloor  \frac{N-1}{2}\right\rfloor  $. The matrix $C\left(  V\right)  $ also
contains information about approximating the eigenvectors of $A$ by points on
$S^{N-1}$: indeed $\min\left\{  \sum_{j=1}^{N}\left\vert \psi_{j}%
-cV_{jk}\right\vert ^{2}:\left\vert c\right\vert =1,\psi\in S^{N-1}\right\}
=2-\sqrt{2\left(  1+C\left(  V\right)  _{kk}^{1/2}\right)  }$. In particular if some
$\lambda_{k}$ is an extreme point in the convex hull of $\left\{  \lambda
_{j}:1\leq j\leq N\right\}  $ then $\lambda_{k}$ is in the real numerical
range of $A$ if and only if $\left\vert \sum_{j=1}^{N}V_{jk}^{2}\right\vert
=1$. This is illustrated by Fig.~\ref{fig:unitary5real}. 


\section{Product numerical shadow} 
Consider the shadow restricted to the set of pure product states. More formally,
we assume that $\mathcal{H}_D = \mathcal{H}_N \otimes \mathcal{H}_M$, apply the
definition (\ref{rshadow}) and take ${\cal R}=\{|\psi_A\rangle \otimes |
\psi_B\rangle \}$, where $|\psi_A \rangle \in \mathcal{H}_N$, while
$|\psi_B\rangle \in \mathcal{H}_M $ and these states are normalized. The set
$\cal R$ of separable (product) pure states has the structure of the Cartesian
product ${\cal R}=\CP{N-1} \times \CP{M-1} \subset \CP{MN-1}$.

The simplest case of $N=M=2$ corresponds to the two--qubit case. The set of
separable (product) pure states has then a form of the Cartesian product of two
spheres ${\cal S}=\Sphere{2} \times \Sphere{2}$. In other words this set forms a
\emph{Segre embedding}, $\CP{1} \times \CP{1} \subset \CP{3}$.

\begin{figure}[h]
\begin{center}
\setlength{\wdth}{0.29\textwidth}
\subfigure[\ $B_{4a}$]{
\includegraphics[width=\wdth]{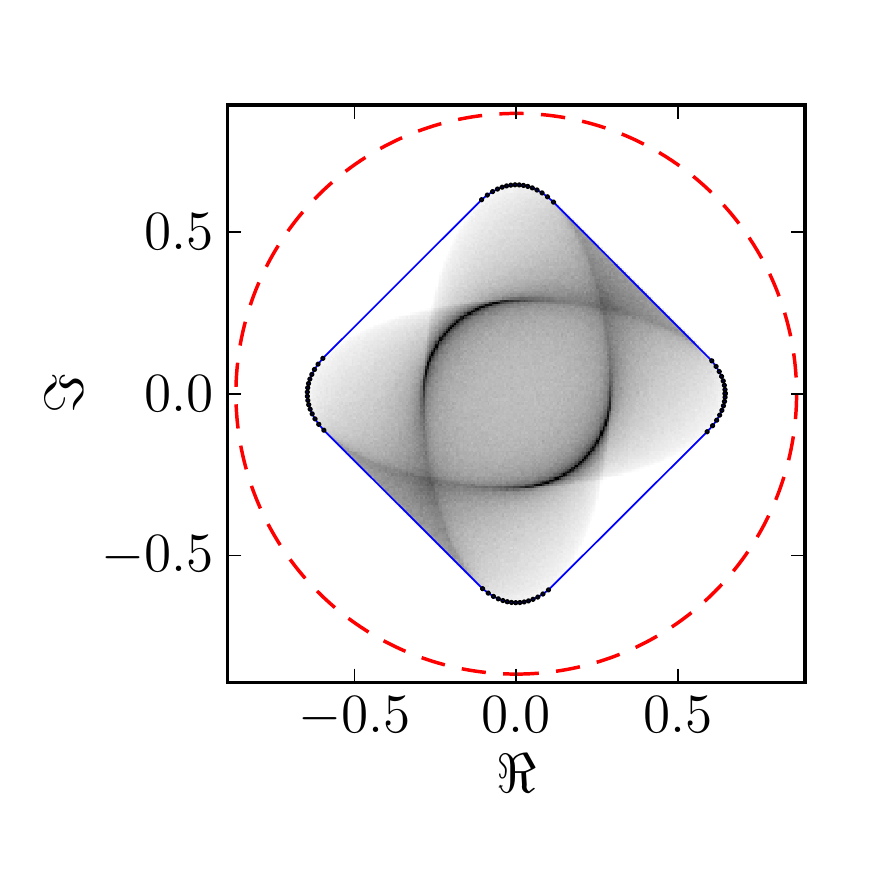}
\label{fig:complex_separable_SA46}
}
\subfigure[\ $B_{4b}$]{
\includegraphics[width=\wdth]{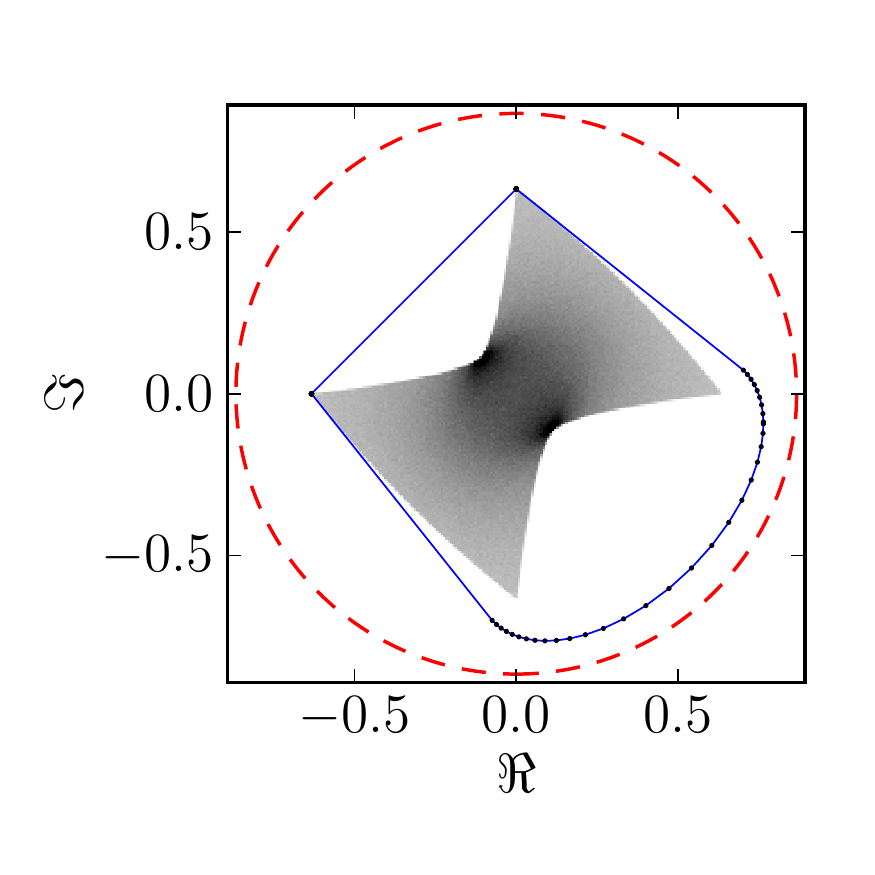}
\label{fig:complex_separable_SA44}
}
\subfigure[\ $B_{4c}$]{
\includegraphics[width=\wdth]{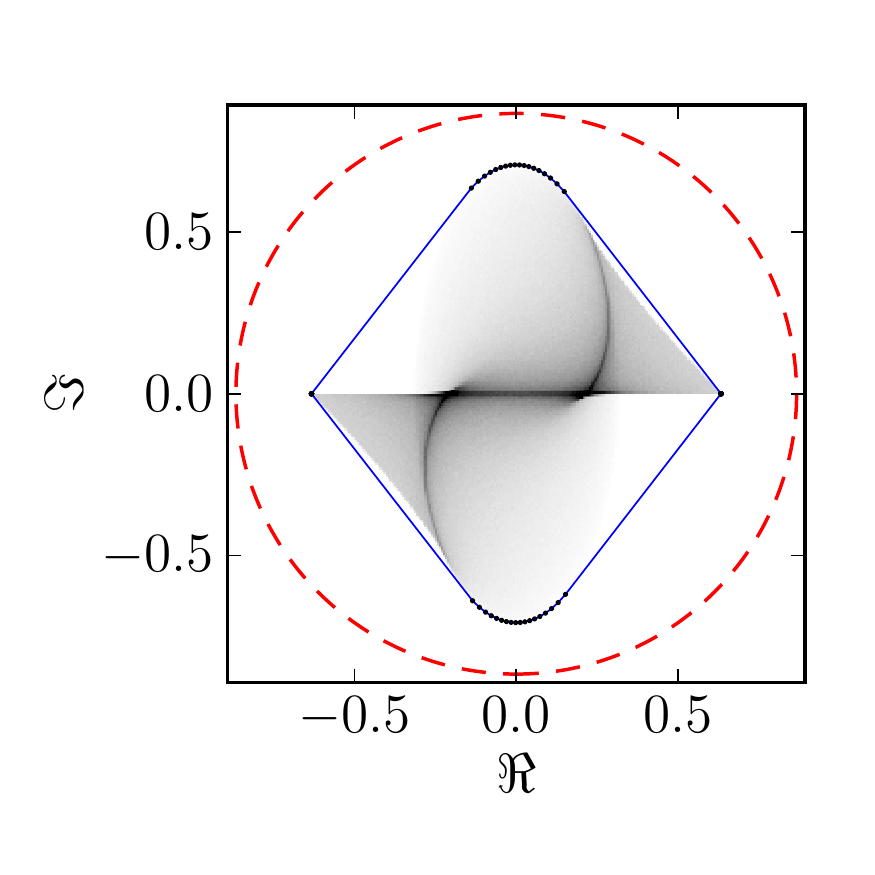}
\label{fig:complex_separable_SA47}
}
\subfigure[\ $B_{4d}$]{
\includegraphics[width=\wdth]{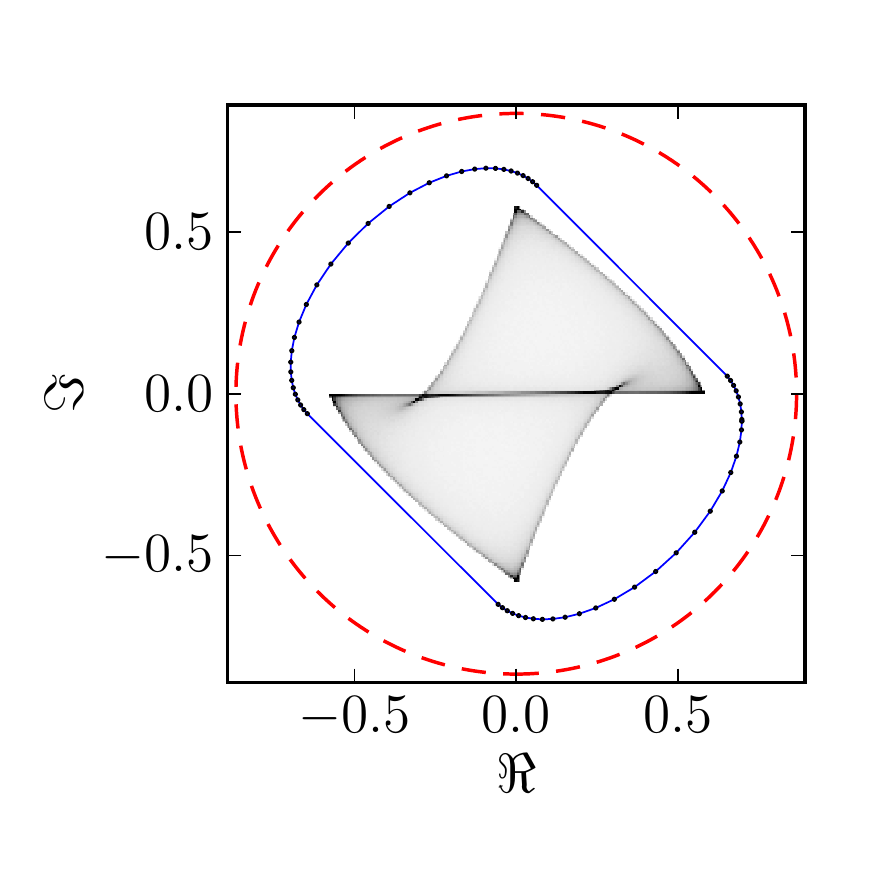}
\label{fig:real_separable_SA43}
}
\subfigure[\ $B_{4e}$]{
\includegraphics[width=\wdth]{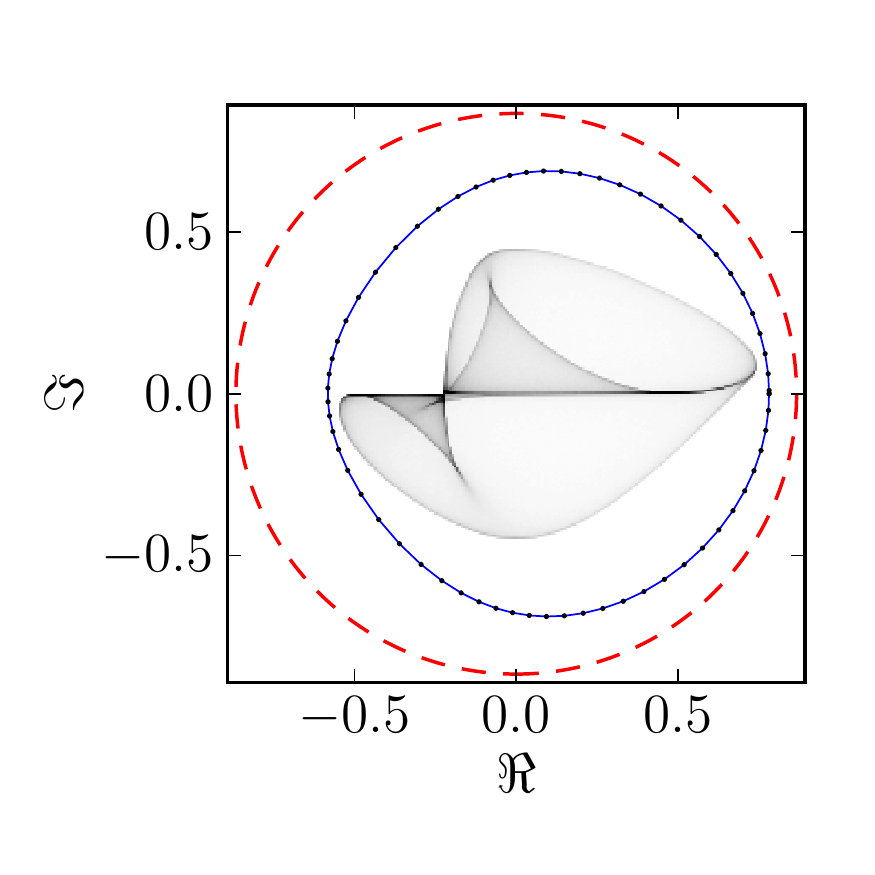}
\label{fig:real_separable_SA40}
}
\subfigure[\ $B_{4f}$]{
\includegraphics[width=\wdth]{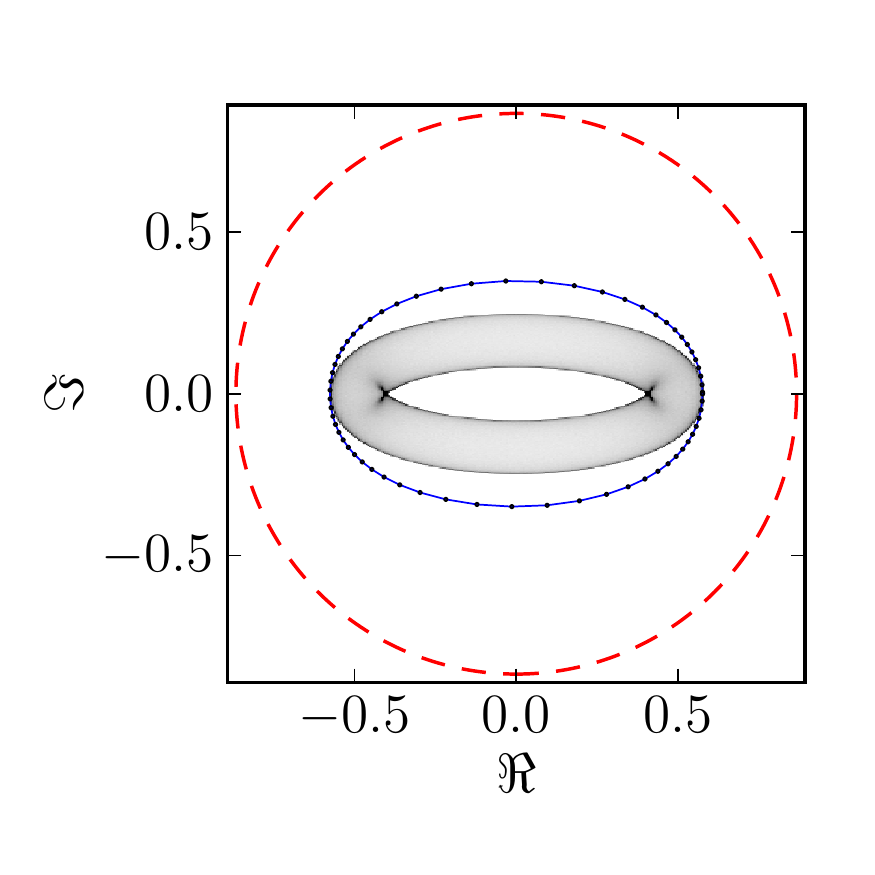}
\label{fig:real_separable_S3B41}
}
\caption{Product numerical shadow for illustrative operators of size $N=4$ with
respect to complex separable states, panels 
\subref{fig:complex_separable_SA46},
\subref{fig:complex_separable_SA44}
and
\subref{fig:complex_separable_SA47},
form projections of
$\CP{1} \times \CP{1} \subset \CP{3}$, while projections with respect to  real
separable states shown in panels 
\subref{fig:real_separable_SA43},
\subref{fig:real_separable_SA40} and
\subref{fig:real_separable_S3B41}
present projections of  $\RP{1}
\times \RP{1}$  equivalent to the torus $T^2$. 
Dashed circle represents the 
image of the outsphere while dotted line denotes the standard numerical range.
Plots are done for matrices translated in such a way
that their traces are equal to zero and suitably rescaled as described in \cite{DGHMPZ11}.
} \label{fig:shadsep}
\end{center}
\end{figure}

One may also consider the shadow with respect to real separable states. In the
two-qubit case this shadow corresponds to a projection of the product of real
projective spaces $\RP{1} \times \RP{1} \subset \RP{3}$, which forms a torus
$\Sphere{1} \times \Sphere{1}=T^2$. Such a structure can be recognized on some
plots shown in Fig. \ref{fig:shadsep} \subref{fig:real_separable_SA43}--
\subref{fig:real_separable_S3B41}.

The matrices used to obtain projections are following
\begin{eqnarray*}
B_{4a} = 
\left[
\begin{array}{cccc}
 1 & 0 & 1 & 0 \\
 0 & \ii & 0 & 1 \\
 0 & 0 & -1 & 0 \\
 0 & 0 & 0 & -\ii
\end{array}
\right],
\quad
&
B_{4b} = 
\left[
\begin{array}{cccc}
 1 & 0 & 0 & 1 \\
 0 & \ii & 0 & 0 \\
 0 & 0 & -1 & 0 \\
 0 & 0 & 0 & -\ii
\end{array}
\right],
\quad\\
B_{4c} = 
\left[
\begin{array}{cccc}
 1 & 0 & 0 & 0 \\
 0 & \ii & 0 & 1 \\
 0 & 0 & -1 & 0 \\
 0 & 0 & 0 & -\ii
\end{array}
\right]
&
B_{4d} = 
\left[
\begin{array}{cccc}
 1 & 0 & 0 & 1 \\
 0 & \ii & 1 & 0 \\
 0 & 0 & -1 & 0 \\
 0 & 0 & 0 & -\ii
\end{array}
\right],
\quad\\
B_{4e} = 
\left[
\begin{array}{cccc}
 1 & 1 & 1 & 1 \\
 0 & \ii & 1 & 1 \\
 0 & 0 & -1 & 1 \\
 0 & 0 & 0 & -\ii
\end{array}
\right],
\quad&
B_{4f} = 
\left[\begin{array}{cc}
\ii&\frac{1}{2}\\
0&\frac{1}{2}\ii
\end{array}\right]\otimes \1 + \1 \otimes 
\left[\begin{array}{cc}
0&2\\
1&\ii
\end{array}
\right].
\end{eqnarray*}

One can note that the numerical ranges presented in Fig.~\ref{fig:shadsep} have a
particular structure. Matrix $B_{4a}$ is permutation equivalent to a simple sum
of two matrices so its numerical range forms a convex hull of two ellipses one
with focal points $\{1,-1\}$ and the other $\{\ii,-\ii\}$. As these ellipses do intersect
their convex hull contains four interval segments. A similar situation occurs
for the matrix $B_{4d}$ where two ellipses -- one with focal points $\{1,-\ii\}$ and
the other $\{\ii,-1\}$ -- do not intersect and the convex hull has only two flat
lines. For the matrix $B_{4b}$ the numerical range is a convex hull of an
ellipse with focal points at eigenvalues $\{1,-\ii\}$ and the line segment between two eigenvalues
$\{\ii,-1\}$. A similar situation arises for the matrix $B_{4c}$.

Consider a particular case of an operator $X$ with the 
 tensor product structure, $X=A\otimes B$. Then
its product numerical range is equal to the Minkowski product of numerical
ranges $\Lambda^{\otimes}(A\otimes B)= \Lambda(A) {\boxtimes} \Lambda(B)$ -  for
more information see \cite{GPMSCZ09}.

Is in this case the shadow of $A\otimes B$ restricted to product states
can be expressed by the numerical shadows of both operators,
\begin{eqnarray}
\mu^{\otimes}_{A\otimes B}  (E) &=& 
\int_{{\mathbbm C \setminus \{0\}}} \mu_{B}\left(\frac{E}{t}\right)  d \mu_A(t) + 1_{\{0\in E\}} \mu_A(\{0\}).
\end{eqnarray}
Here $\mu_A$ and $\mu_B$ denote the probability measures related to the
numerical shadows of $A$ and $B$ respectively, while $E \subset {\mathbbm C}$ is
a measurable subset of the complex plane.

\subsection{Mean and variance for the separable numerical shadow}
In the case of separable shadow of matrix $X \in \mathcal{M}_{N^2}$ it is
possible to obtain explicit expressions for the mean and the variance. We have 
\begin{equation} 
\mathbb{E}\left(  z\right)  =\frac{1}{N^{2}}\mathrm{tr}X \; ,
\label{mean-sep-shad}
\end{equation}
and 
\begin{eqnarray} 
\mathbb{E}\left(  z\overline{z}\right) & = & 
\frac{1}{N^2 (N+1)^2} 
 \left(
|\tr X|^2 
+ \|\tr_A X\|_{\mathrm{HS}}^2 
+ \|\tr_B X\|_{\mathrm{HS}}^2 
+ \|X\|_{\mathrm{HS}}^2 
\right).
\label{var-sep-shad}
\end{eqnarray}
Formulas involve the partial traces ${\rm tr}_A X$ and ${\rm tr}_B X$,
follows from more general fact given in Appendix \ref{appen:2}.
These expressions imply directly the following result. 
\begin{equation} \label{var-sep}
\mathrm{Var}(z) = \frac{1}{N^2 (N+1)^2} 
 \left(
 - \frac{2 N + 1}{ N^2} |\tr X|^2 
+ \|\tr_A X\|_{\mathrm{HS}}^2 
+ \|\tr_B X\|_{\mathrm{HS}}^2 
+ \|X\|_{\mathrm{HS}}^2 
\right).
\end{equation}

\subsection{Separable numerical shadow for a diagonal matrix}

Consider a diagonal matrix $X$ defined on a composite Hilbert space, ${\cal
H}_D={\cal H}_N \otimes {\cal H}_M$ of dimension $D=NM$. Its diagonal elements
forming the spectrum $\{x_i\}_{i=1}^D$ can be also represented by two indices,
$\{y_{\mu \nu} \}$ with $\mu =1,\dots,N$ and $\nu=1,\dots,M$.

Let $|1,1\rangle$ be an arbitrary fixed pure product state in ${\cal H}_D$, so
the set of random separable pure states can be obtained as $|\psi\rangle = U
|1,1\rangle= W\otimes V |1,1\rangle$, where $W\in \mathrm{U}(N)$ and $V\in
\mathrm{U}(M)$ are independent random unitary matrices distributed according to
the Haar measure. Thus the expansion coefficients of the product state
$|\psi\rangle$ read $(U_{11}, U_{12},\dots, U_{1D})=(W_{11} V_{11},\dots, W_{1N}
V_{1M})$.

The \emph{separable numerical shadow} of  the diagonal operator $X$ is defined as
the density distribution of random numbers $z:=\langle \psi|X|\psi\rangle$, 
where $|\psi\rangle$ is a separable random state defined by random unitaries 
$U$ and $V$. In this case one has 

\begin{equation}
z:=\langle 1,1|(W^{\dagger}\otimes V^{\dagger})  X (W \otimes V) |1,1\rangle = 
\sum_{i=1}^D x_i |U_{1i}|^2 =
\sum_{\mu=1}^N \sum_{\nu=1}^M y_{\mu \nu}  |W_{1\mu}|^2 |V_{1\nu}|^2 =
x\cdot r
\label{prod1}
\end{equation}
where $r$ is a real probability vector of size $D=NM$.
It can be considered as a tensor product of 
two probability vectors $p \in \Delta_{N-1}$ and $q \in \Delta_{M-1}$,
since its components read $r_{\mu \nu}=p_{\mu} q_{\nu}$ with 
with $\mu =1,\dots N$ and $\nu=1,\dots,M$.

Thus the separable numerical shadow of a diagonal operator can be considered as
a projection of the \emph{Cartesian product} of classical probability simplices,
$ \Delta_{N-1} \times \Delta_{M-1}$. In the simplest case of $D=4=2\times 2$ the
Cartesian product of two intervals (1-simplices) forms a square, which lives
inside the tetrahedron of the $4$--dimensional probability vectors.

As in the previous case we can distinguish two probability measures
in the space of unitary matrices. They lead to 
\begin{itemize}
    \item[C)] \emph{complex separable shadow}, generated by the Haar measure on
    $\mathrm{U}(N)$ and $\mathrm{U}(M)$, for which both probability vectors $p$
    and $q$ are distributed uniformly with respect to the Lebesgue measure on
    the simplices $\Delta_{N-1}$ and $\Delta_{M-1}$, respectively.
    \item [D)] \emph{Real separable shadow}, generated by the Haar measure on the
    orthogonal groups $\mathrm{O}(N)$ and $\mathrm{O}(M)$, which lead to the
    statistical measure (Dirichlet measure with $s=1/2$) in both simplices.
\end{itemize}

Note that in these case the separable shadow of $X$ is supported on its \emph{
product numerical range} \cite{GPMSCZ09}, which in general forms a proper subset
of the convex hull of the spectrum. The product structure of the classical
probability vector $r$ in (\ref{prod1}), generalized for a multiple tensor
product structure, is consistent with the parametrization of the product
numerical range described in Prop. 12 in \cite{GPMSCZ09}.

\section{Maximally entangled numerical shadow}

Consider an operator $X$ acting on a Hilbert space with a tensor product
structure, ${\cal H}={\cal H}_A \otimes {\cal H}_B$. For simplicity let us
assume that the dimensions of both subspaces are equal to $N$ so the total
dimension reads $D=N^2$. Among all pure states of the $N \times N$ system one
distinguishes the set $\cal E$ of maximally entangled states. It contains the
states equivalent with respect to a local unitary operation $U_A \otimes U_B$ to
the generalized Bell state, $|\psi_+\rangle=\frac{1}{\sqrt{N}} \sum_i
|i,i\rangle$. Thus the set of maximally entangled states has the structure of
$\mathrm{U}(N)/\mathrm{U}(1)=\mathrm{SU}(N)/Z_N$, where $Z_N$ is the discrete
permutation group \cite{SZK02}, \cite[Ch. 15]{BZ06}. Choosing ${\cal E}$ for the
set $R$ in (\ref{rshadow}) we define the shadow $P^{\cal E}_X(z)$ of an operator
$X$ with respect to the maximally entangled states. The corresponding
probability measure will be denoted as $d \mu^{\cal E}_X(z)$.

\subsection{Two qubit case: $D= 2\times 2$}

\begin{figure}[ht!]
\begin{center}
\setlength{\wdth}{0.3\textwidth}
\subfigure[\ $B_{4b}$]{
\includegraphics[width=\wdth]{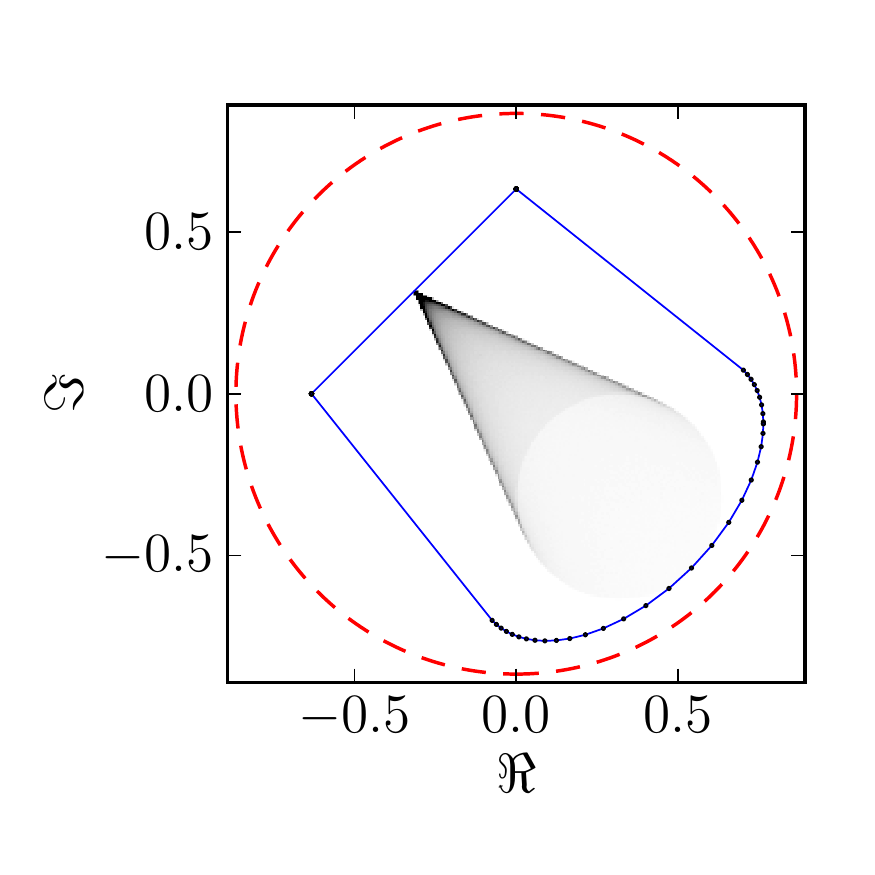}
\label{fig:complex_entangled_SA44}
}
\subfigure[\ $B_{4e}$]{
\includegraphics[width=\wdth]{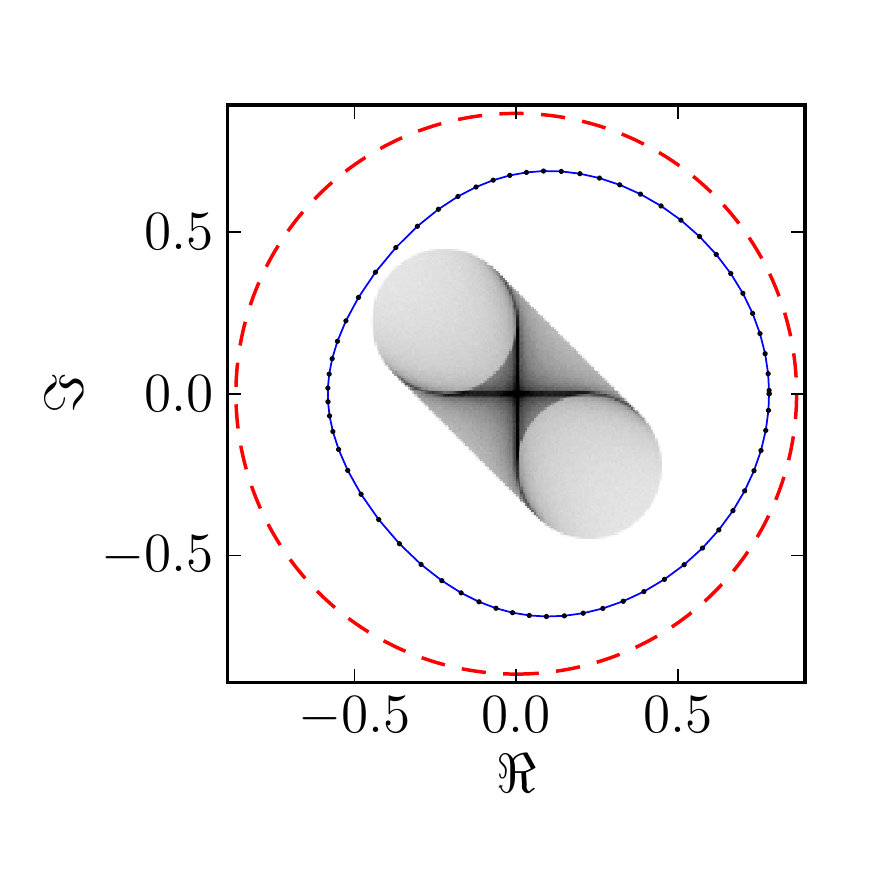}
\label{fig:complex_entangled_SA40}
}
\subfigure[\ $B_{4c}$]{
\includegraphics[width=\wdth]{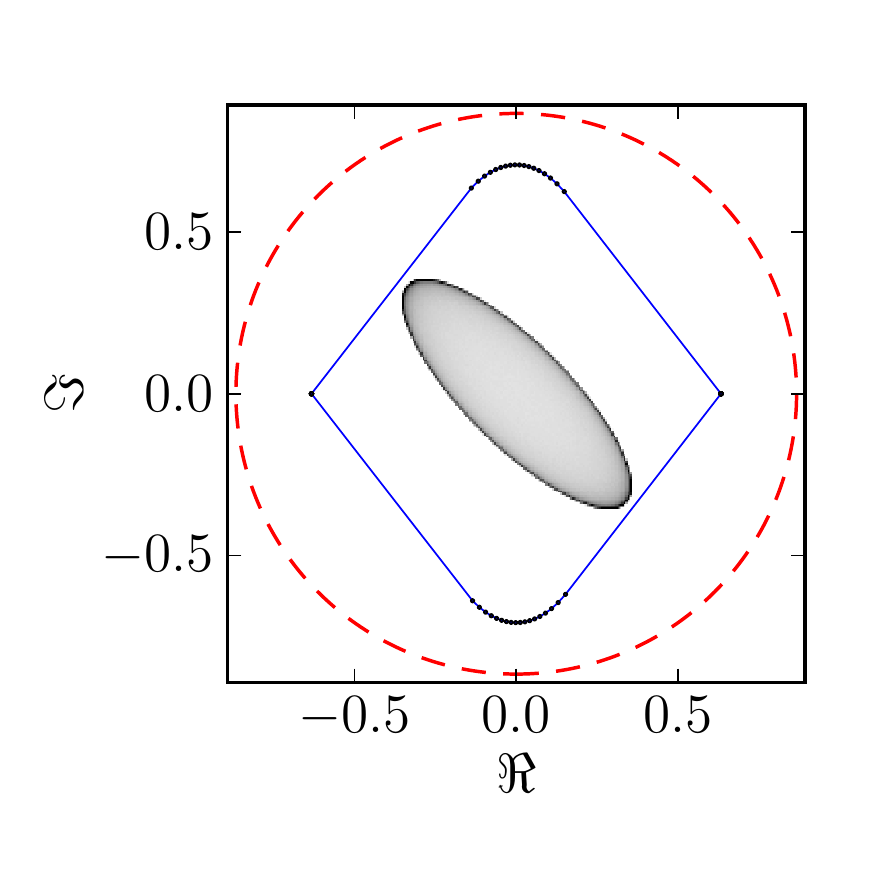}
\label{fig:complex_entangled_SA47}
}
\subfigure[\ $B_{4b}$]{
\includegraphics[width=\wdth]{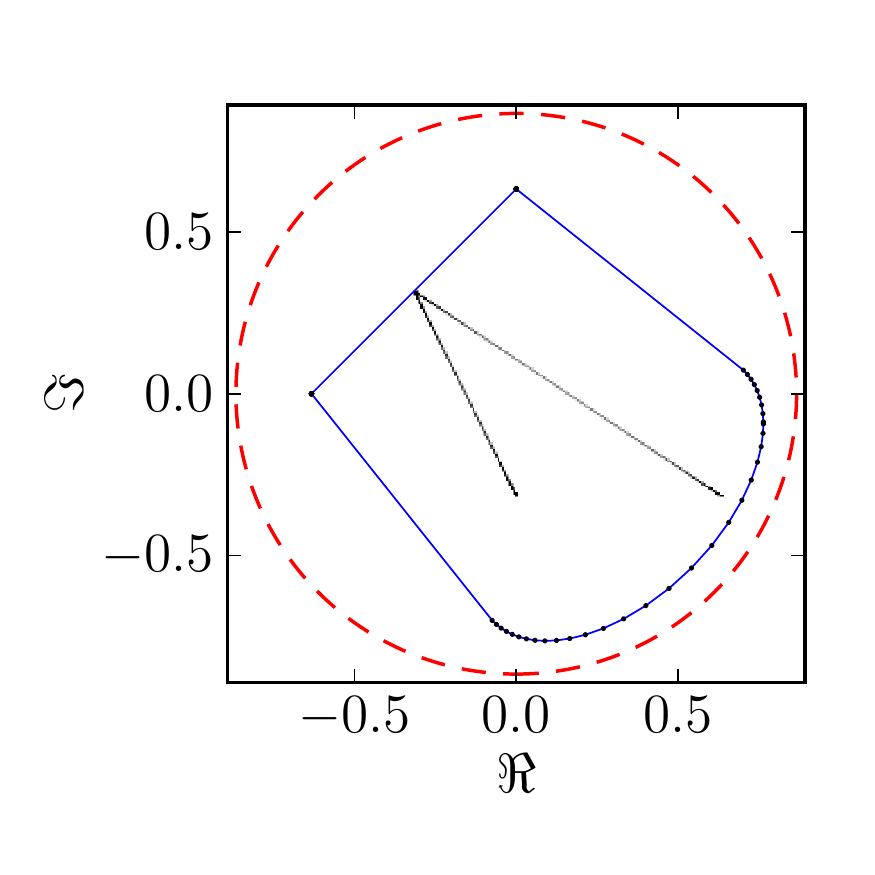}
\label{fig:real_entangled_SA44}
}
\subfigure[\ $B_{4e}$]{
\includegraphics[width=\wdth]{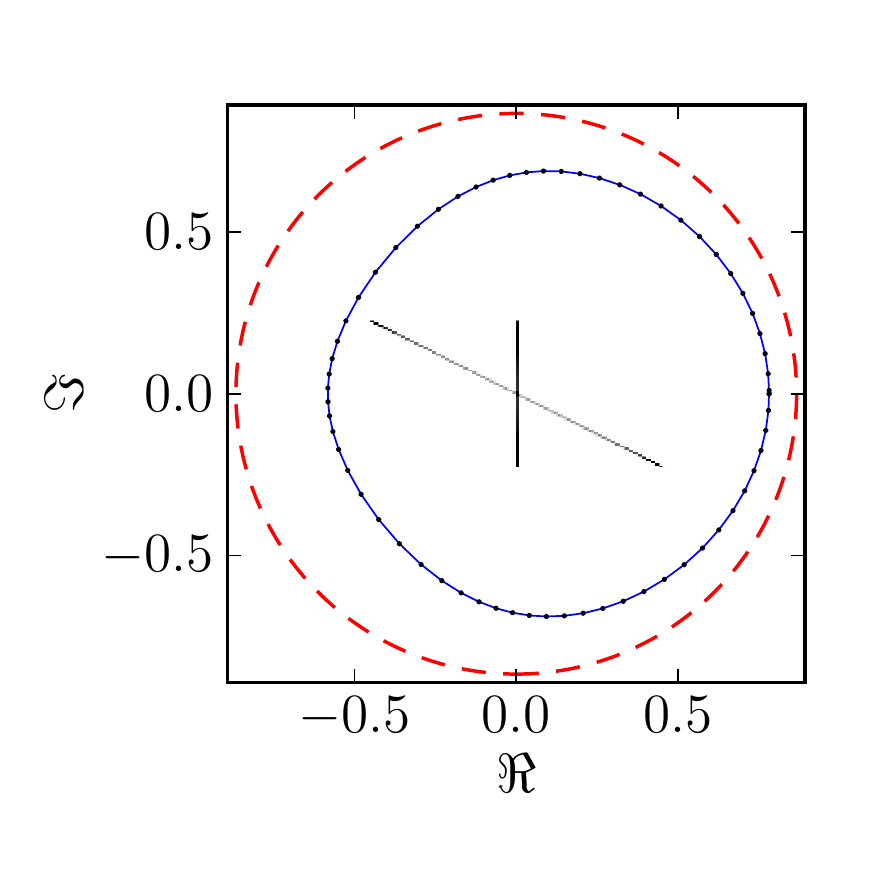}
\label{fig:real_entangled_SA40}
}
\subfigure[\ $B_{4c}$]{
\includegraphics[width=\wdth]{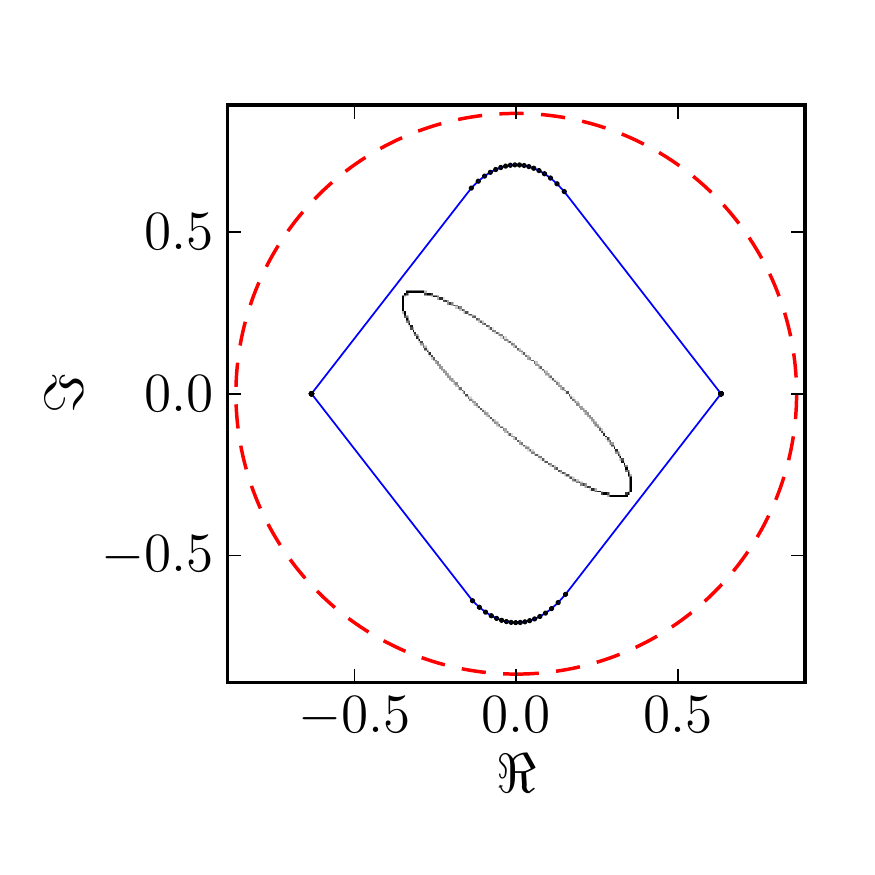}
\label{fig:real_entangled_SA47}
}
\caption{Entangled numerical shadow for illustrative operators of size $N=4$
with respect to complex maximally entangled states, panels 
\subref{fig:complex_entangled_SA44}--\subref{fig:complex_entangled_SA47},
form  projections of $\mathrm{U}(2)/\mathrm{U}(1)\sim \RP{3}$,
while projections with respect to  real entangled states 
shown in panels 
\subref{fig:real_entangled_SA44}--\subref{fig:real_entangled_SA47}
present projections of $\mathrm{O}(2)/\mathrm{U}(1)$ equivalent to a circle $\Sphere{1}$.
Plots are done for matrices translated in such a way
that their traces are equal to zero and suitably rescaled as described in \cite{DGHMPZ11}.
}
\label{fig:shadent}
\end{center}
\end{figure}

In the simplest case of $2 \times 2$ Hilbert space 
the set  $\cal E$ of maximally entangled states
has the structure $\mathrm{U}(2)/\mathrm{U}(1)=\RP{3}$ --- see \cite{BZ06}. 
Hence the numerical shadow of an operator $A$ of order four
with respect to the complex maximally entangled states
can be considered as a projection of the real 
projective space on the plane --- see  the shadow for
some illustrative operators presented in Fig. \ref{fig:shadent} 
\subref{fig:complex_entangled_SA44}--\subref{fig:complex_entangled_SA47}.

If one considers a further restriction
and studies the shadow with respect to \emph{real} 
maximally entangled states, the result can be interpreted 
as an image of the space $\mathrm{O}(2)/\mathrm{O}(1)=\RP{1} = \Sphere{1}$ 
Observe that the illustrative shadows obtained in this case
and presented in Fig. \ref{fig:shadent} 
\subref{fig:real_entangled_SA44}--\subref{fig:real_entangled_SA47}
show indeed projections of a circle onto the complex plane.
 
In the special case of a diagonal operator $B$ of size four
its shadow with respect to complex maximally entangled states
can be identified with a standard shadow of a reduced operator $B'$  
of size $2$. This fact is formulated in the following proposition,
proved in Appendix \ref{appen:1}.

\begin{proposition}
\label{prop:2x2ent}
Consider an diagonal matrix of order four, $X={\rm diag} (d_1, d_2, d_3, d_4)$
which acts on a composite Hilbert space ${\cal H}={\cal H}_A \otimes {\cal H}_B$
and the reduced matrix $Y=Y(X)= \frac{1}{2} {\rm diag} (d_1+d_4, d_2+d_3)$. Then
the numerical shadow $P^{\cal E}_X(z)$ of $X$ with respect to complex maximally
entangled states $\cal E$ is equal to the standard numerical shadow $P_Y(z)$ of
the reduced matrix $Y(X)$ of order two.
\end{proposition}

\subsection{Two quNit case: $D= N\times N$}

In this general case the set $\cal E$ of maximally entangled states forms a
manifold of $(N^2-1)$ real dimensions with the structure of $\mathrm{U}(N)/\mathrm{U}(1)$
\cite{BZ06}. In the case of the shadow with respect to real maximally entangled
states the space under consideration reads $\mathrm{O}(N)/\mathrm{O}(1)$. 

To analyse the entangled shadow of a diagonal matrix
we can perform initial steps used to prove proposition \ref{prop:2x2ent}.
It will be convenient to use a slightly different notation
and consider a diagonal matrix $X$ of size $D=N^2$ acting on 
a tensor product space ${\cal H}={\cal H}_A \otimes {\cal H}_B$
with entries
$X_{\left(  i,j\right)  ,\left(  i,j\right)  }$
where $i,j=1,\dots,N$.
Consider a local unitary matrix $U^{(A)} \otimes U^{(B)}$ where $U^{(A)}, U^{(B)} \in \mathrm{U}(N)$.
%
The entangled unit state $v$ has entries
\begin{equation}
v_{\left(  i,j\right)  }   =\frac{1}{\sqrt{N}}\sum_{k=1}^{N}
U_{ik}^{(A)} U_{jk}^{(B)} 
 =\frac{1}{\sqrt{N}} U_{ij}
%
\end{equation}
where $U=U^{(A)} \left( U^{(B)}\right)^T$.
For purposes of numerical shadow the integration with respect to the Haar measure
over both matrices $U^{(A)}$ and $U^{(B)}$
can be replaced by a single integration over the $N$ 
dimensional random unitary matrix $U$.
Observe $U\mapsto
\overline{U}$ is a real automorphism of $\mathrm{U}\left( N\right)  $ 
invariant for the Haar measure.

Taking a diagonal matrix 
$X_{\left(  i_{1},j_{1}\right)
,\left(  i_{2},j_{2}\right)  }=\delta\left(  i_{1},i_{2}\right)  \delta\left(
j_{1},j_{2}\right)  C_{i_{1},j_{1}}$
 we find its expectation value for random pure state $v$
\[
v^{\dag}Av=\frac{1}{N}\sum_{i,j=1}^{N} C_{ij}\left\vert U_{ij}\right\vert
^{2}=\frac{1}{N}\mathrm{tr}\left( CB^{T}\right)  .
\]
Here we reshape the diagonal matrix $X$ of order $N^2$
to get a matrix $C$ of order $N$ with entries $C_{ij}$,
while  $B$ stands for a unistochastic matrix,
 $B_{ij}:= \vert U_{ij}\vert^{2},$ for $i,j=1,\dots, N$. 
 The  case $N=2$ studied above relied on the
simple nature of the set of unistochastic matrices of order two, 
 equivalent to an interval. It is known that the
structure of unistochastic matrices for $N\ge 3$ is complicated
and interesting \cite{BEKTZ05,DZ09}. Thus
we are not able to formulate a direct generalization of proposition 
\ref{prop:2x2ent} for the $N \times N$ problem.

For the case of $3\times3$ matrices consider the function%
\[
v^{\dag}Av=\frac{1}{3} \tr\left(  CB^{T}\right)  ,
\]
where the (variable) unistochastic matrix reads%
\[
B=\left[
\begin{array}
[c]{ccc}%
b_{1} & b_{2} & 1-b_{1}-b_{2}\\
b_{3} & b_{4} & 1-b_{3}-b_{4}\\
1-b_{1}-b_{3} & 1-b_{2}-b_{4} & b_{1}+b_{2}+b_{3}+b_{4}-1
\end{array}
\right]  .
\]
Given the matrix $C$ let $C_{i\cdot}=\sum_{k=1}^{3}C_{ik}$, $C_{\cdot j}%
=\sum_{k=1}^{3}C_{kj},$ for $1\leq i,j\leq3$, and $C_{\cdot\cdot}=\sum
_{i,j=1}^{3}C_{ij}$. For simplification we set%
\[
\gamma_{ij}=C_{ij}-\frac{1}{3}C_{i\cdot}-\frac{1}{3}C_{\cdot j}+\frac{1}%
{9}C_{\cdot\cdot} \ .%
\]
One can observe, that the row and column sums of $\left[\gamma_{ij}\right]$
are zero and
\begin{align*}
v^{\dag}Av  &  =\frac{1}{3}tr\left[
\begin{array}
[c]{cc}%
4b_{1}+2b_{2}+2b_{3}+b_{4}-3 & 2b_{1}+4b_{2}+b_{3}+2b_{4}-3\\
2b_{1}+b_{2}+4b_{3}+2b_{4}-3 & b_{1}+2b_{2}+2b_{3}+4b_{4}-3
\end{array}
\right]  \left[
\begin{array}
[c]{cc}%
\gamma_{11} & \gamma_{12}\\
\gamma_{21} & \gamma_{22}%
\end{array}
\right] +\frac{1}{9}C_{\cdot\cdot} \ .
\end{align*}
Without loss of generality we can assume $C_{\cdot\cdot}=0$. Powers of
$v^{\dag}Av$ can be integrated using the formula from \cite[Prop.~3.3]{Du1}:%
\begin{align*}
&  \int_{\mathrm{U}\left(  3\right)  }b_{1}^{n_{1}}b_{2}^{n_{2}}b_{3}^{n_{3}}%
b_{4}^{n_{4}}dm\\
&  =\frac{n_{1}!n_{2}!n_{3}!n_{4}!\left(  2\right)  _{n_{2}+n_{3}}\left(
2\right)  _{n_{1}+n_{2}+n_{4}}}{\left(  3\right)  _{n_{1}+n_{2}+n_{3}+n_{4}%
}\left(  2\right)  _{n_{2}+n_{4}}\left(  2\right)  _{n_{1}+n_{2}}\left(
2\right)  _{n_{3}}}~_{4}F_{3}\left(
\genfrac{}{}{0pt}{}{-n_{1},-n_{2},-n_{4},1+n_{3}}{1,2+n_{3},-1-n_{1}%
-n_{2}-n_{4}}%
;1\right),
\end{align*}
where $dm$ denotes the Haar measure on $\mathrm{U}(3)$. 

For example%
\[
\int_{\mathrm{U}\left(  3\right)  }\left(  v^{\dag}Av\right)  ^{2}dm=\frac{1}%
{72}\left\{  3\sum_{i,j=1}^{2}\gamma_{ij}^{2}+\left(  \sum_{i,j=1}^{2}%
\gamma_{ij}\right)  ^{2}+2\left(  \gamma_{11}+\gamma_{22}\right)  \left(
\gamma_{12}+\gamma_{21}\right)  \right\}  .
\]

There is an interesting special case when one of the variables $b_{i}$, say
$b_{4}$, does not appear in $v^{\dag}Av$. The triple $\left(  b_{1}%
,b_{2},b_{3}\right)  $ is a point in the pyramid with square base $\left\{
\left(  0,b_{2},b_{3}\right)  :0\leq b_{2},b_{3}\leq1\right\}  $ and vertex
$\left(  1,0,0\right)$. The induced measure (from $\mathrm{U}\left(  3\right)  $) on
the pyramid is $\frac{2}{1-b_{1}}db_{1}~db_{2}~db_{3}$. The numerical range of
$A$ is an affine image of the pyramid, hence is the convex hull of the images
of the vertices of the pyramid, $\left\{  \left(  1,0,0\right)
,\left(  0,1,0\right)  ,\left(  0,0,1\right)  ,\left(  0,1,1\right)  ,\left(
0,0,0\right)  \right\}  $, that is, a convex polygon. For three arbitrary
complex numbers $z_{1},z_{2},z_{3}$ let%
\[
\left[
\begin{array}
[c]{cc}%
\gamma_{11} & \gamma_{12}\\
\gamma_{21} & \gamma_{22}%
\end{array}
\right]  =\left[
\begin{array}
[c]{cc}%
-6z_{1}-6z_{2} & 3z_{1}+3z_{3}\\
3z_{2}+3z_{3} & -3z_{3}%
\end{array}
\right]  ,
\]
then $v^{\dag}Av=\left(  z_{3}-2z_{1}-2z_{2}\right)  b_{1}+\left(  z_{3}%
-z_{1}\right)  b_{2}+\left(  z_{3}-z_{2}\right)  b_{3}+z_{1}+z_{2}-z_{3}$. The
vertices of the pyramid are mapped to $\left\{  z_{1},z_{2},z_{3},-z_{1}%
-z_{2},z_{1}+z_{2}-z_{3}\right\}  $. It is possible that these points form a
(not regular) pentagon. For example let $z_{1}=1,z_{2}=\ii,z_{3}=-\frac{3}%
{4}+\frac{\ii}{8}$ then the range is the convex hull of $\left\{  1,\frac{7}%
{4}+\frac{7}{8}\ii,\ii,-\frac{3}{4}+\frac{\ii}{8},-1-\ii\right\}  $.

It may not be easy to find the shadow measure explicitly, but one expects a
higher density in the neighbourhood of $-z_{1}-z_{2}$, the image of $b=\left(
1,0,0\right)$. Fix $\varepsilon>0$ and consider the set $B_{\varepsilon
}=\left\{  \left(  b_{1},b_{2},b_{3}\right)  :1-\varepsilon<b_{1}%
<1,0<b_{2},b_{3}<1-b_{1}\right\}  $. The normalized volume of $B_{\varepsilon
}$ is $\varepsilon^{3}$ and the $\mathrm{U}\left(3\right)$-measure of
$B_{\varepsilon}$ is $\varepsilon^{2}$, so the relative density is $\frac
{1}{\varepsilon}$.

\subsection{Mean and variance for the entangled shadow}

It is possible to get explicit expressions for the mean and the variance of the
entangled shadow $d\mu^{\cal E}_{X}$ of matrix $X$ acting on $N \times N$
Hilbert space ${\cal H}={\cal H}_A \otimes {\cal H}_B$. The following results
\begin{equation} 
\mathbb{E}\left(z\right)=\int_{{\mathbbm C}}z \; d\mu^{\cal E}_{X}\left(z\right)
=\frac{1}{N^{2}} \mathrm{tr}X \; ,
\label{meaneshad}
\end{equation}
and 
\begin{eqnarray} 
\mathbb{E}\left(  z\overline{z}\right) &=& 
\int_{{\mathbbm C}}z\overline{z}\; d\mu^{\cal E}_{X} \left(z\right)  =\frac{1}{N^{2}
\left(  N^{2}-1\right)  }
\left\{ \|X\|_{\mathrm{HS}}^2 + 
|\tr X |^2 \right\} \nonumber  \\
& - & \frac{1}{N^{3}\left(  N^{2}-1\right)  }
\left\{
\|\tr_A X\|_{\mathrm{HS}}^2 + \|\tr_B X\|_{\mathrm{HS}}^2 
\right\},
\label{varentshad}
\end{eqnarray}
which involve the partial traces ${\rm tr}_A X$ and ${\rm tr}_B X$,
are derived in Appendix \ref{appen:2} from more general fact.
These expressions imply directly the following result. 

\medskip
\begin{proposition}
The expected squared distance from the mean 
 with respect to the  entangled  shadow $d\mu^{\cal E}_{X}$ reads
\begin{eqnarray} 
\int_{{\mathbbm C}}\left\vert z-\mathbb{E}\left(  z\right)  \right\vert ^{2} 
d\mu^{\cal E}_{X}\left(  z\right) &=&
\frac{1}{N^{2}\left(  N^{2}-1\right)  }
\left\{
 \|X\|_{\mathrm{HS}}^2 
+
\frac{1}{N^{2}} |\tr X|^2 
\right\} \nonumber \\
& -& 
\frac{1}{N^{3}\left(  N^{2}-1\right)  }
\left\{
\|\tr_A X\|_{\mathrm{HS}}^2 + \|\tr_B X\|_{\mathrm{HS}}^2  
\right\}  . 
\label{distestshad}
\end{eqnarray}
\end{proposition}
\medskip

Let us apply these formulae in the special case of a diagonal matrix 
$X_{\left(  i_{1},j_{1}\right) ,\left(  i_{2},j_{2}\right)  }=\delta\left(  i_{1},i_{2}\right)  \delta\left(
j_{1},j_{2}\right)  C_{i_{1},j_{1}}$.
In this case the necessary ingredients of (\ref{distestshad})
simplify considerably, e.g.
$$
\mathrm{tr}X   =\sum_{i,j=1}^{N } C_{ij}, \quad \quad
\mathrm{tr}\left(  XX^{\dag}\right)
=\sum_{i,j=1}^{N}\left\vert C_{ij}\right\vert ^{2},$$
$$ \left(  \mathrm{tr}_A X\right) _{j_{1},j_{2}}  
=\delta\left( j_{1},j_{2}\right) \sum_{i=1}^{N} C_{i,j_{1}},\quad \quad
\left(  \mathrm{tr}_B X\right) _{i_{1},i_{2}}  
=\delta\left( i_{1},i_{2}\right) \sum_{j=1}^{N} C_{i_{1},j}, $$
$$ \mathrm{tr}\left(  \left(  \mathrm{tr}_A X\right)  \left(
\mathrm{tr}_A X^{\dag}\right)  \right)    =\sum_{j=1}^{N}\left\vert \sum_{i=1}^{N}C_{ij}\right\vert ^{2},\quad \quad 
\mathrm{tr}\left(  \left(  \mathrm{tr}_B X\right)  \left(
\mathrm{tr}_B X^{\dag}\right)  \right)   =\sum_{i=1}^{N}\left\vert \sum_{j=1}^{N}C_{ij}\right\vert ^{2}.
$$

For the easy case $N=2$ we find $\mathbb{E}\left(  \left\vert z-\mathbb{E}\left(  z\right)
\right\vert ^{2}\right)  =\frac{1}{48}\left\vert C_{11}+C_{22}-C_{12}%
-C_{21}\right\vert ^{2}$, in agreement with the previous calculations.

\section{Shadow and dynamics of quantum entanglement}

In previous sections we analysed the entire set of quantum states with its
subsets and their projections onto a plane. In this section we specify a
concrete quantum dynamics (in general non--unitary), choose an initial quantum
state $\rho(0)$, and following \cite{DGHMPZ11} we analyse its trajectory
projected on the plane of a shadow of a selected non-Hermitian operator $A$. In
particular we will be interested in dynamics of quantum entanglement, so the
separable shadow of $A$ will be used as a background for the trajectory obtained
from the expectation values, $z(t)= {\rm Tr} A \rho(t)$.

Investigation of the dynamics of quantum entanglement was initiated in
\cite{ZHHH01}, in which the evolution of certain measures of entanglement in
time was studied for a model non unitary dynamics of a two qubit system and
several qualitatively different scenarios of behaviour of entanglement in time
were identified. In particular, revivals of entanglement in time and an effect
of sudden decay of quantum entanglement was reported. The latter effect was
later called \emph{entanglement sudden death} by Yu and Eberly \cite{YE04,YE09}
and the dynamics of entanglement was studied by several authors in various
setups \cite{MKZ05,FT06,DJ06,DL08,FT08}.

\begin{figure}[ht]
\setlength{\wdth}{0.42\textwidth}
\centering
\subfigure[\ Schematic figure often used to explain
           dynamics of quantum entanglement]{
\includegraphics[width=1.6\wdth]{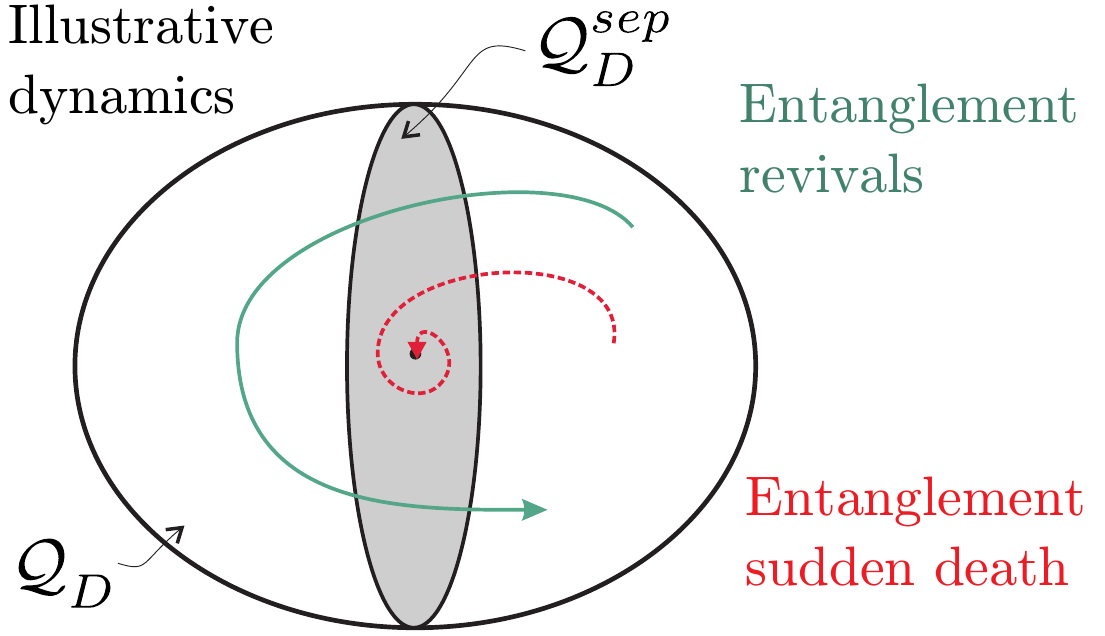}
\label{fig:dynentaglement-concept}
}\\
\centering
\subfigure[\ Actual trajectory observed with the
separable shadow of an exemplary matrix $X_1$ in the background. 
All points outside this shadow are entangled 
(red crosses along the trajectory),
while the points projected into the centre 
of the separable shadow are separable
(blue circles along the trajectory).]{
\includegraphics[width=\wdth]{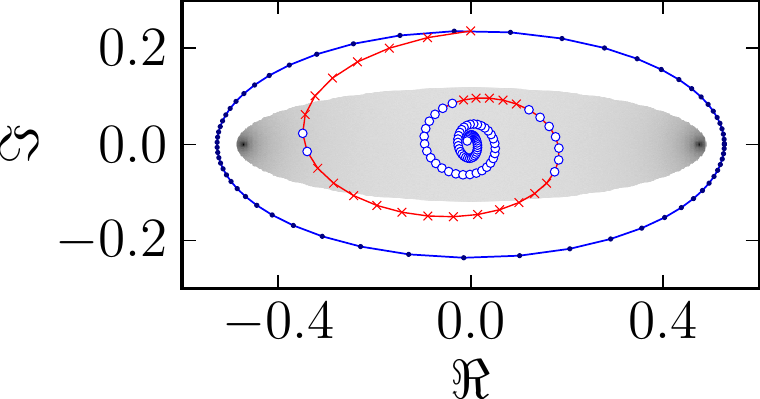}
\label{fig:dynentaglement-1}
}\quad
\centering
\subfigure[\  The same trajectory observed by matrix $X_2$. 
States of the trajectory 
projected into the separable shadow are
typically  separable.]{
\includegraphics[width=\wdth]{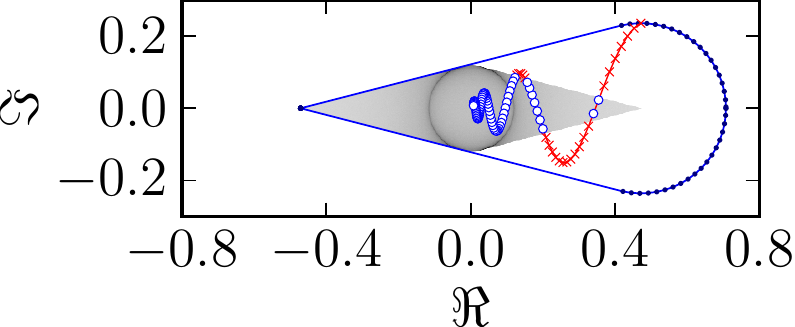}
\label{fig:dynentaglement-2}
}
\caption{Dynamics of quantum entanglement.}
\label{fig:dynentaglement}
\end{figure}

In general, all these dynamical effects can be explained in a simple geometric
manner, if one takes into account the structure of the set of separable states
analysed in \cite{KZ01}. As the convex set ${\cal Q}^{\rm sep}_D$ of separable
mixed states of a bipartite system occupies the central part of the set ${\cal
Q}_D$ of all states of size $D$, entanglement revivals occur if the unitary
dynamics moves the initial state several times across the separability boundary.
On the other hand, entanglement sudden death effect takes place, if the
decoherence is so strong that the initially pure state gets mixed in such a pace
that it crosses the separability boundary only once. These possible scenarios
are shown on a schematic sketch, used in a conference talks for several years
--- see Fig. \ref{fig:dynentaglement-concept}.

Making use of the technique of numerical shadow we are now in position to
observe similar behaviour of entanglement for a concrete choice of quantum
dynamics and initial states. Consider the following discrete-time dynamic
quantum process of a system consisting of two qubits \cite{ZHHH01}. The system
is assumed to be initially described by a maximally entangled pure state,
$\rho_0=\frac{1}{2}(\ket{00}+\ket{11})(\bra{00}+\bra{11})$. The discrete time
evolution of the system is given by one--step unitary evolution expressed by
Pauli matrices, $U=e^{\ii \alpha \sigma_x\otimes\sigma_y}$, followed by an
action of the depolarising channel acting locally on the second qubit,
\begin{equation}
\Phi(\rho)=
\sqrt{(1-\beta)}\1_4\, \rho\, \sqrt{(1-\beta)}\1_4+
\sum_{p\in\{x,y,z\}}
\sqrt{\frac13\beta}\1_2\otimes\sigma_p\, 
\rho\, \sqrt{\frac13\beta}\1_2\otimes\sigma_p,
\end{equation}
so that $\rho_{t+1}=U\Phi(\rho_t)U^\dagger$.

As the initial state is chosen to be pure, the trajectory begins at the boundary
of the set of mixed states. As time $t$ increases the trajectory plunges into
the set of mixed states periodically crossing the set of entangled states.
Artists impression of such processes is depicted in
Fig.~\ref{fig:dynentaglement-concept} where the outer oval corresponds to the
set of the pure states, its interior corresponds to the set of all mixed states,
ellipse corresponds to the set of separable states and spirals depict the
trajectory.

Let us fix the parameters of the discussed process, by setting the interaction
strength $\alpha=0.1$ and decoherence rate $\beta=0.03$. We arbitrarily chose
two matrices
\begin{equation}
X_1=
\left[
\begin{array}{cccc}
 -1 & 0 & 0 & \ii \\
 0 & 1 & 0 & 0 \\
 0 & 0 & -1 & 0 \\
 0 & 0 & 0 & 1 \\
\end{array}
\right]
\text{ and }
X_2=
\left[
\begin{array}{cccc}
 1 & 0 & 0 & \ii \\
 0 & -1 & 0 & 0 \\
 0 & 0 & -1 & 0 \\
 0 & 0 & 0 & 1 \\
\end{array}
\right],
\end{equation}
which allow us to project the trajectory $\rho_t$ onto the complex plane. We 
can calculate the images of the trajectory $z_t^{(k)}=\tr(\rho_t^\dagger X_k), 
k=1,2$ and superimpose them on the separable shadows of matrices $X_k$. The 
resulting images are shown in Fig.~\ref{fig:dynentaglement} panels 
\subref{fig:dynentaglement-1} and
\subref{fig:dynentaglement-2}. 
The red crosses indicate entangled states and blue circles indicate separable 
states. 

It can be easily seen that by choosing an appropriate observation matrix it is 
convenient to observe the dynamics of entanglement in the process.
In the general case for any given trajectory, it is hardly 
possible to find such 
an observation matrix whose product numerical range contains only images of 
separable states, but images of entangled states always lie outside of product 
numerical range. Therefore product numerical range and separable shadow of a 
matrix are useful tools to visualize dynamics of entanglement, the effects of 
entanglement sudden death and entanglement revival.

\section{Multipartite systems} 
It is natural to ask about the properties of the numerical shadow in the case
when one aims to study composite quantum systems consisting of more than two
subsystems.

Let us consider the simplest case of a multipartite quantum system, \emph{i.e.}
a system composed of three qubits. In this case $N=2^3=8$.
As an example we take an unitary matrix $U_8$ of size eight
written in the standard computational basis
$\{|0,0,0\rangle, |0,0,1\rangle, \dots , |1,1,1\rangle \}$
\begin{equation}
\label{eqn:thomas-nonsimply-example}
U_8 ={\rm diag}
(1,
e^{\frac{2 \ii \pi }{3}},
e^{\frac{2 \ii \pi }{3}},
e^{-\frac{2 \ii \pi }{3}},
e^{\frac{2 \ii \pi }{3}},
e^{-\frac{2 \ii \pi }{3}},
e^{-\frac{2 \ii \pi }{3}},
1).
\end{equation}

\begin{figure}[ht]
\centering
\setlength{\wdth}{0.3\textwidth}
\subfigure[\ Standard shadow]{
\includegraphics[width=\wdth]{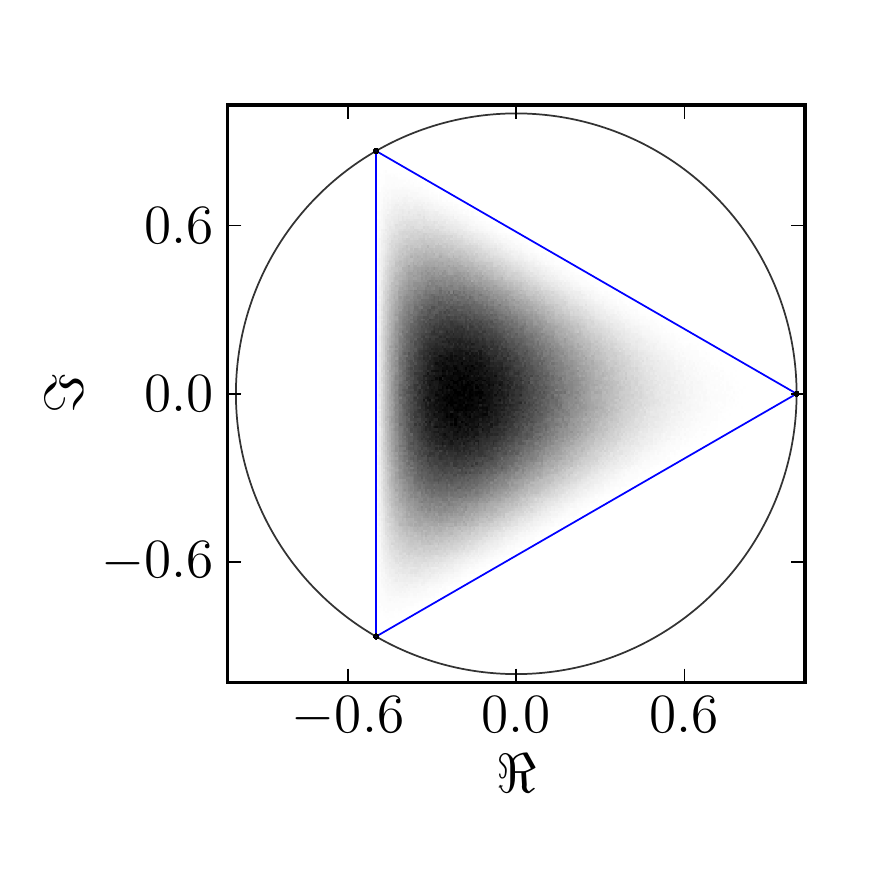}
\label{fig:thomas-standard}
}
\subfigure[\ Separable shadow]{
\includegraphics[width=\wdth]{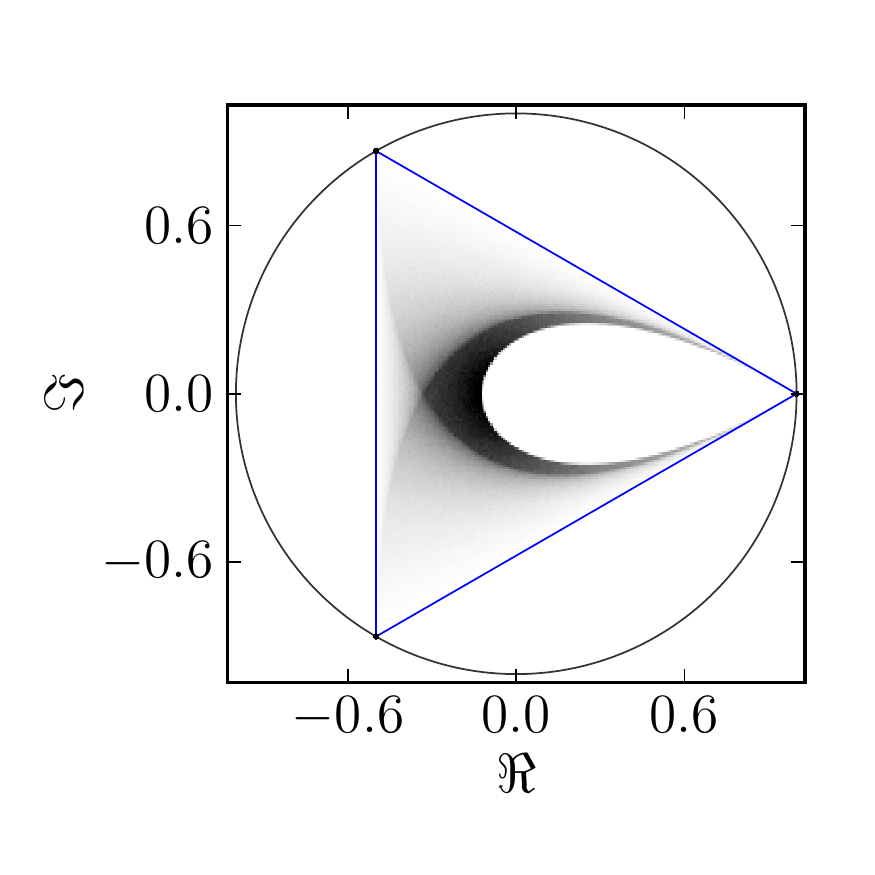}
\label{fig:thomas-separable}
}\\
\subfigure[\ GHZ--entangled states]{
\includegraphics[width=\wdth]{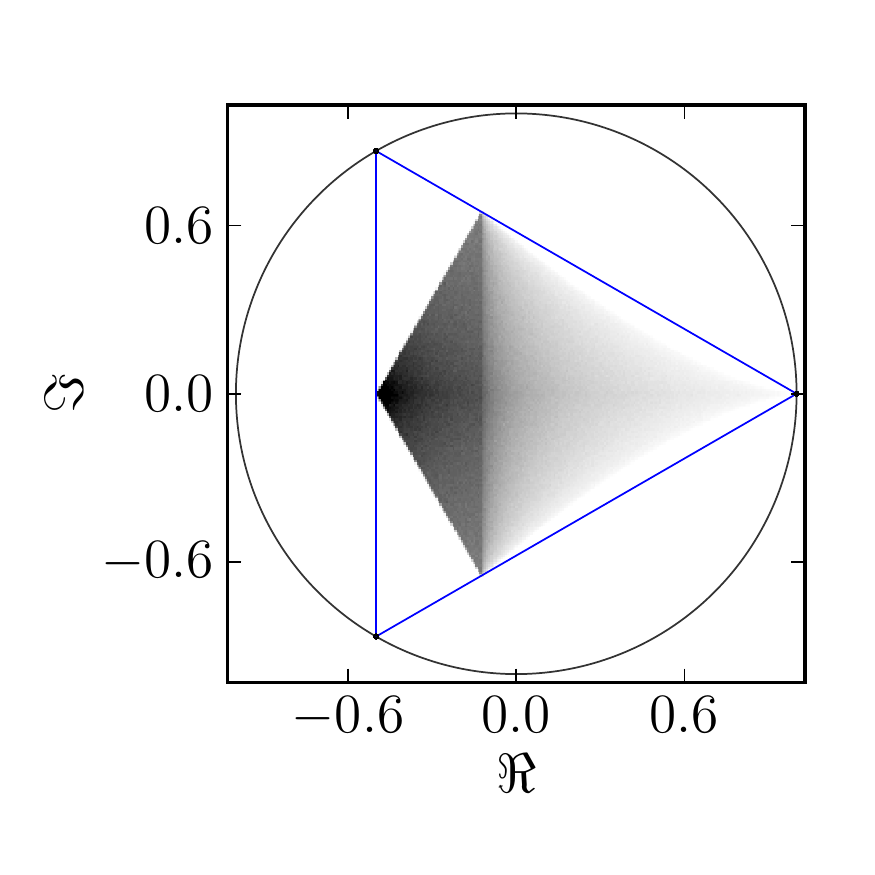}
\label{fig:thomas-ghz}
}
\subfigure[\ W--entangled states]{
\includegraphics[width=\wdth]{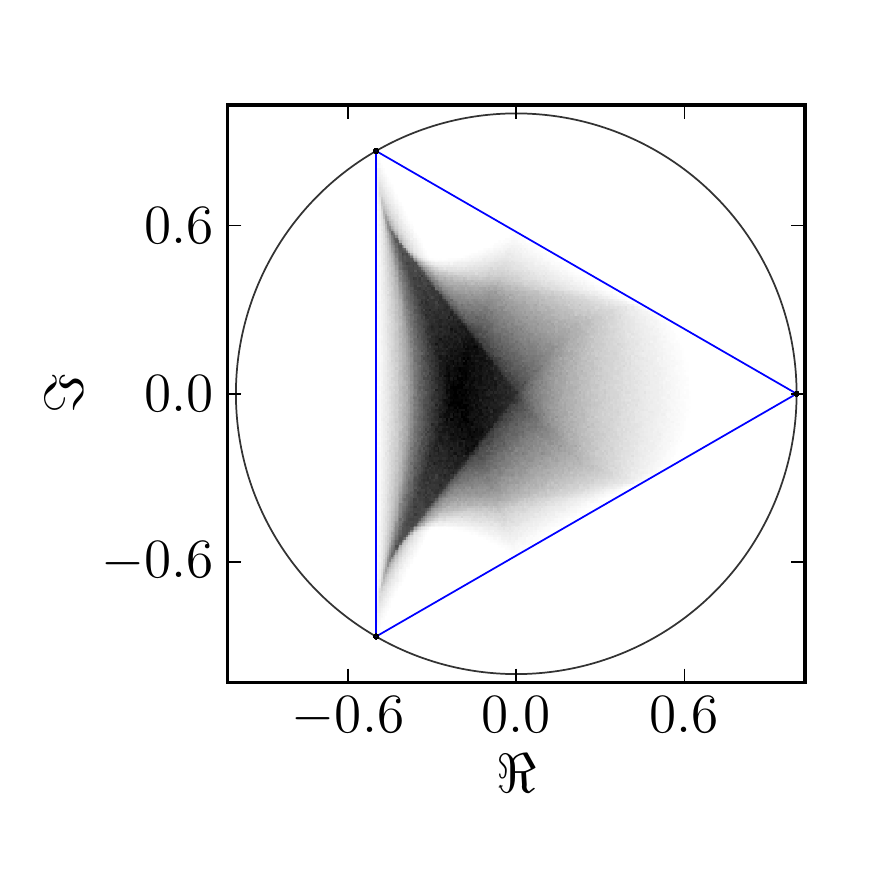}
\label{fig:thomas-W}
}
\caption{Restricted shadows given by matrix from
Eq.~\ref{eqn:thomas-nonsimply-example}. In each case the probability
distribution is supported on a subset of a numerical range, namely restricted
numerical range. In the general case this subset is not convex and the
restricted numerical shadow can be supported on non convex set.}
\label{fig:thomas-nonsimply-example}
\end{figure}

The product numerical range of this operator is not simply connected
\cite{SHDHG08,GPMSCZ09}, so it is instructive to study
 the shadow of $U_8$ with respect to the space of

\newcounter{typeofstate}
\begin{list}{\alph{typeofstate})}{\usecounter{typeofstate}}
	\item all pure states (standard shadow);
	
	\item product states (product shadow), $|\psi\rangle_{\rm sep}=U_A\otimes
	U_B \otimes U_C |0,0,0\rangle$;

	\item GHZ entangled states, $|\psi\rangle_{\rm GHZ}=U_A\otimes U_B \otimes
	U_C (|0,0,0\rangle +|1,1,1\rangle)/\sqrt{2}$;

	\item so-called $W$--entangled states, $|\psi\rangle_{\rm W}=U_A\otimes U_B
	\otimes U_C (|1,0,0\rangle +|0,1,0\rangle+|0,0,1\rangle)/\sqrt{3}$.
\end{list}
Here $U_A, U_B, U_C$ are independent random unitary matrices taken from $\mathrm{U}(2)$
with respect to the Haar measure. Fig.~\ref{fig:thomas-nonsimply-example}
presents the shadows generated by matrix $U_8$ with respect to those classes of
states.

Observe that a generic operator acting on the three-qubit system leads to
different shadows, if they are taken with respect to $GHZ$ states and the $W$
states. This is a consequence of the different topology of the orbits with
respect to local unitary transformations produced by these two classes of
entangled states. Study of numerical shadows restricted to certain classes of
entangled states can contribute to a better understanding of the geometry of the
manifold of locally equivalent states.

Investigations of the restricted shadows of operators acting on multipartite
systems lead to a wide class of interesting problems which in general are
difficult to solve. However, some results obtained in the previous sections for
the bipartite setup can be generalized for multipartite systems.

Consider an operator acting on the composite Hilbert space describing
$m$--partite system of dimensions $N_1,N_2, \dots, N_m$, respectively. Its
shadow with respect to separable states $|\Psi\rangle=|\psi_1\rangle \otimes
|\psi_2\rangle \otimes \dots |\psi_m\rangle$ leads to classical product measures
on a simplex of a composed dimension $D=N_1 N_2 \cdots N_m-1$, induced by the
Dirichlet measures on $m$ simplices of the dimension $(N_i-1)$ with $i=1,\dots ,
m$.

\section{Concluding remarks}

In this work we propose to combine the notion of restricted numerical range with
the numerical shadow of an operator which is a probability measure on the
complex plane. On one hand the numerical shadow can be investigated for a given
operator $X$. On the other hand, one may analyse the shadows of all operators of
a fixed dimension $D$, which can be considered as projections of the set of all
pure states of size $D$ onto a plane.

In a similar way it could be interesting to study the shadow of a given operator
$X$ with respect to various sets of pure states. For instance, in this work, we
analysed the standard shadow with respect to complex states and the shadow
restricted to real states. For operators acting on composite Hilbert spaces one
can additionally study the shadow with respect to separable or maximally
entangled states, complex or real. Note that in general the restricted numerical
range is not convex \cite{DHKSH08,PGMSCZ10}, which implies that the restricted
numerical shadow can be supported on non convex sets.

Following the complementary strategy one may take the set of all operators of a
given dimension and analyse their shadows restricted to a certain class of
states. These probability distributions on the complex plane convey some
information about the structure of these specific subsets of the set of all
quantum states. Consider for instance the simplest Hilbert space with a tensor
product structure ${\cal H}_2 \otimes {\cal H}_2$. Then the standard shadow
carries information about the set of all pure states $\Omega_4$ which forms the
complex projective space $\CP{3}$, while the shadow restricted to real states
corresponds to the real projective space $\RP{3}$.

In a similar way, the shadows of matrices of order four with respect to
separable states illustrate the projections of the product of two spheres
$\CP{1} \times \CP{1} = \Sphere{2} \times \Sphere{2}$, while the real separable
shadow corresponds to projections of a torus $\RP{1} \times \RP{1} =\Sphere{1}
\times \Sphere{1} = T^2$. The shadow with respect to maximally entangled states
(also called briefly the \emph{entangled shadow}) represents the set ${\cal
E}=\mathrm{U}(2)/\mathrm{U}(1)$ also equivalent to real projective space
$\RP{3}$. In the case of real entangled shadow we observe projections of the set
${\cal E}_R=\mathrm{O}(2)/\mathrm{O}(1)$ equivalent to the circle $\RP{1} \sim
\Sphere{1}$.

The notion of separable shadow is useful to analyse the dynamics of quantum
entanglement. For a given initial quantum state $\rho(0)$ and a certain dynamics
we select a non-Hermitian matrix $A$ and study trajectories on the complex plane
formed by the time evolution of its expectation value, $z(t)={\rm Tr} A\rho(t)$.
Investigations of such a trajectory superimposed on the separable shadow of $A$
contribute to our understanding of the dynamics of quantum entanglement and
allow us to visualize the effects of entanglement sudden death and entanglement
revivals.

The notion of restricted numerical range is easily formulated for operators
acting on Hilbert space with multiple tensor product structure \cite{PGMSCZ10},
which correspond to multipartite quantum systems. Therefore it is natural to
define restricted numerical shadow for various classes of quantum states of
multipartite systems. For instance, in the simplest case of a three-qubit
system, described in the Hilbert space ${\cal H}_8={\cal H}_2^{\otimes 3}$ one
may distinguish two classes of maximally entangled states called $GHZ$ and $W$,
which cannot be locally converted in any direction \cite{DVC00}. Studying
numerical shadows of matrices of order $8$, restricted to $GHZ$ states or $W$
states, we are thus in position to investigate the differences between the
structure of these two important classes of three-qubit quantum entangled
states.

In conclusion, we have introduced the notion of the restricted numerical shadow
of an operator and established its basic properties. On one hand we believe that
this topic is interesting from the mathematical point of view, as it relates
operator theory and probability. Moreover, we are tempted to expect that further
investigations of the restricted numerical shadow will contribute to a better
understanding of the geometry of quantum entanglement, so that they become
directly applicable to the theory of quantum information.

\subsection*{acknowledgements}
It is a pleasure to thank E. Gutkin for fruitful discussions. Work by
J.~Holbrook was supported in part by an NSERC of Canada research grant. Work by
P.~Gawron was supported by the Polish National Science Centre under the grant
number N N516 481840, Z.~Pucha{\l}a was supported by Polish National Science
Centre under the research project N N514 513340, J.A.~Miszczak was supported by
Polish Ministry of Science and Higher Education under the research project
IP2011 036371, while K.~\.Zyczkowski acknowledges support by the Polish Ministry
of Science and Higher Education grant number N202 090239. Numerical calculations
presented in this work were performed on the \texttt{Leming} server of The
Institute of Theoretical and Applied Informatics, Polish Academy of Sciences.

\appendix

\section{Proof of Proposition \ref{prop:2x2ent}}\label{appen:1}
To analyse properties of the shadow with respect to complex maximally entangled
states for an operator acting on the $2 \times 2$ Hilbert space let us analyse
the structure of a local unitary transformation $U^{(A)} \otimes U^{(B)}$ acting
on ${\cal H}={\cal H}_A \otimes {\cal H}_B$. 
Consider two generic elements of $\mathrm{U}\left(2\right)  $
\begin{align*}
U^{\left(  A\right)  }  
  =\left[
\begin{array}
[c]{cc}%
e^{\ii\left(  \psi_{1}+\phi_{1}\right)  }\cos\theta_{1} & -e^{\ii\left(  \psi
_{1}-\phi_{2}\right)  }\sin\theta_{1}\\
e^{\ii\left(  \psi_{1}+\phi_{2}\right)  }\sin\theta_{1} & e^{\ii\left(  \psi
_{1}-\phi_{1}\right)  }\cos\theta_{1}%
\end{array}
\right]  ,\ \ \ 
U^{\left(  B\right)  }  
  =\left[
\begin{array}
[c]{cc}%
e^{\ii\left(  \psi_{2}+\phi_{3}\right)  }\cos\theta_{2} & -e^{\ii\left(  \psi
_{2}-\phi_{4}\right)  }\sin\theta_{2}\\
e^{\ii\left(  \psi_{2}+\phi_{4}\right)  }\sin\theta_{2} & e^{\ii\left(  \psi
_{2}-\phi_{3}\right)  }\cos\theta_{2}%
\end{array}
\right]  ,
\end{align*}
where $-\pi<\phi_{j}\leq\pi,0\leq\psi_{j}<\pi,0\leq\theta_{j}\leq\frac{\pi}%
{2}$ and the Haar measure (for $U^{\left(  A\right)  }$) is%
\[
dm\left(  U\right)  =\frac{1}{2\pi^{3}}d\psi_{1}d\phi_{1}d\phi_{2}\sin
\theta_{1}\cos\theta_{1}d\theta_{1}.
\]
Taking tensor products of corresponding columns of $U^{\left(  A\right)  }$
and $U^{\left(  B\right)  }$ and adding them
 we obtain a parametrization of an entangled state,
\[
v=\frac{1}{\sqrt{2}}\left[
\begin{array}
[c]{c}%
U_{11}^{\left(  A\right)  }U_{11}^{\left(  B\right)  }+U_{12}^{\left(
A\right)  }U_{12}^{\left(  B\right)  }\\
U_{21}^{\left(  1\right)  }U_{11}^{\left(  2\right)  }+U_{22}^{\left(
A\right)  }U_{12}^{\left(  B\right)  }\\
U_{11}^{\left(  1\right)  }U_{21}^{\left(  2\right)  }+U_{12}^{\left(
A\right)  }U_{22}^{\left(  B\right)  }\\
U_{21}^{\left(  1\right)  }U_{21}^{\left(  2\right)  }+U_{22}^{\left(
A\right)  }U_{22}^{\left(  B\right)  }%
\end{array}
\right].
\]

Consider an diagonal matrix of order four,
$X={\rm diag} (d_1, d_2, d_3, d_4)$

so that its expectation value for a maximally entangled state reads
\begin{align*}
v^{\dag}Xv  &  =\frac{1}{2}\left(  d_{1}+d_{4}\right)  q_{1}+\frac{1}%
{2}\left(  d_{2}+d_{3}\right)  q_{2},\\
q_{1}  &  =\cos^{2}\theta_{1}\cos^{2}\theta_{2}+\sin^{2}\theta_{1}\sin
^{2}\theta_{2}+\frac{1}{2}\sin2\theta_{1}\sin2\theta_{2}\cos\left(  \phi
_{1}+\phi_{2}+\phi_{3}+\phi_{4}\right)  ,\\
q_{2}  &  =\cos^{2}\theta_{1}\sin^{2}\theta_{2}+\sin^{2}\theta_{1}\cos
^{2}\theta_{2}-\frac{1}{2}\sin2\theta_{1}\sin2\theta_{2}\cos\left(  \phi
_{1}+\phi_{2}+\phi_{3}+\phi_{4}\right)  .
\end{align*}
Observe that $q_{1}+q_{2}=1$. Compare this situation to the complex shadow of
the reduced operator $Y=Y(X)$
\begin{align*}
Y  & = \left[
\begin{array}
[c]{cc}%
\frac{1}{2}\left(  d_{1}+d_{4}\right)  & 0\\
0 & \frac{1}{2}\left(  d_{2}+d_{3}\right)
\end{array}
\right]  ,\\
x^{\dag}Y x  &  =\frac{1}{2}\left(  d_{1}+d_{4}\right)  \left\vert
x_{1}\right\vert ^{2}+\frac{1}{2}\left(  d_{2}+d_{3}\right)  \left\vert
x_{2}\right\vert ^{2},
\end{align*}
where $x$ belongs to 
   the unit sphere $\Sphere{1}$ in ${\mathbbm C}^{2}$. We know that
\[
\int_{\Sphere{1}}\left\vert x_{1}\right\vert ^{2n}\left\vert x_{2}\right\vert ^{2k}%
d\mu\left(  x\right)  =\frac{n!k!}{\left(  k+n+1\right)  !},~k,n=0,1,2,\ldots
.
\]
To identify the two shadows it only remains to show%
\[
\int_{\mathrm{U}\left(  2\right)  \times \mathrm{U}\left(  2\right)  }q_{1}^{n}q_{2}%
^{k}dm\left(  U^{\left(  1\right)  }\right)  dm\left(  U^{\left(  2\right)
}\right)  =\frac{n!k!}{\left(  n+k+1\right)  !},~k,n=0,1,2,\ldots.
\]
In fact it suffices to show $\int q_{1}^{n}=\frac{1}{n+1}$. The integration
over the angles $\phi_{j}$ can be combined into one $\phi$,
due to the rotational invariance. Write
\begin{align*}
q_{1}  &  =\left(  \cos\theta_{1}\cos\theta_{2}+e^{i\phi}\sin\theta_{1}%
\sin\theta_{2}\right)  \left(  \cos\theta_{1}\cos\theta_{2}+e^{-i\phi}%
\sin\theta_{1}\sin\theta_{2}\right)  ,\\
q_{1}^{n}  &  =\sum_{j=0}^{n}\sum_{l=0}^{n}\binom{n}{j}\binom{n}{l}\left(
\cos\theta_{1}\cos\theta_{2}\right)  ^{2n-j-l}\left(  \sin\theta_{1}\sin
\theta_{2}\right)  ^{j+l}e^{i\left(  j-l\right)  \phi}.
\end{align*}
Now integrate with $\frac{1}{2\pi}d\phi$ over $-\pi<\phi\leq\pi$, then over
the $d\theta_{1}d\theta_{2}$ part:%
\begin{align*}
\int q_{1}^{n}  &  =4\sum_{j=0}^{n}\binom{n}{j}^{2}\int_{0}^{\frac{\pi}{2}%
}\int_{0}^{\frac{\pi}{2}}\left(  \cos\theta_{1}\cos\theta_{2}\right)
^{2n-2j+1}\left(  \sin\theta_{1}\sin\theta_{2}\right)  ^{2j+1}d\theta
_{1}d\theta_{2}\\
&  =\sum_{j=0}^{n}\binom{n}{j}^{2}\left(  \frac{\left(  n-j\right)
!j!}{\left(  n+1\right)  !}\right)  ^{2}=\frac{1}{\left(  n+1\right)  ^{2}%
}\sum_{j=0}^{n}1=\frac{1}{n+1}.
\end{align*}
To complete the proof%
\begin{align*}
\int q_{1}^{n}q_{2}^{k}  &  =\int q_{1}^{n}\left(  1-q_{1}\right)  ^{k}%
=\sum_{j=0}^{k}\binom{k}{j}\left(  -1\right)  ^{j}\int q_{1}^{n+j}\\
&  =\sum_{j=0}^{k}\binom{k}{j}\left(  -1\right)  ^{j}\frac{1}{n+j+1}=\frac
{1}{n+1}\sum_{j=0}^{k}\frac{\left(  -k\right)  _{j}\left(  n+1\right)  _{j}%
}{j!\left(  n+2\right)  _{j}}\\
&  =\frac{\left(  n+2-n-1\right)  _{k}}{\left(  n+1\right)  \left(
n+2\right)  _{k}}=\frac{n!k!}{\left(  n+k+1\right)  !},
\end{align*}
where $(x)_j$ is Pochhammer symbol.
This used the Chu-Vandermonde sum (terminating $_{2}F_{1}\left(  1\right)  $)
and $\dfrac{1}{n+1+j}=\dfrac{\left(  n+1\right)  _{j}}{\left(  n+1\right)
_{j}\left(  n+1+j\right)  }=\dfrac{\left(  n+1\right)  _{j}}{\left(
n+1\right)  \left(  n+2\right)  _{j}}$.

\section{Expectation values for shadows}\label{appen:2}

Consider a general matrix $X$ acting on a $N \times N$ composite Hilbert space
${\cal H}={\cal H}_A \otimes {\cal H}_B$ with entries written in  a four index
notation $X_{\stackidx{i_1}{i_2}{j_1}{j_2}}$, where $i_1,j_1,i_2,j_2=1,\dots,N$.
Upper pair of indices determines the row of the matrix, while the lower pair
determines its column. We use the standard operations on matrices, which in this
notation read 
\begin{equation}
X_{\stackidx{i_1}{i_2}{j_1}{j_2}}^{\dag} := \overline{X}_{\stackidx{j_1}{j_2}{i_1}{i_2}}, 
\quad \quad
\mathrm{tr}X   :=
\sum_{i,j=1}^{N}
X_{\stackidx{i}{j}{i}{j}}
\label{tracex}
\end{equation}
and introduce reduced matrices of size $N$ obtained by a  partial trace over a
single subsystem, 
\begin{equation} 
\left(  \mathrm{tr}_A  X\right)_{i_{2} j_{2}}  
:=\sum_{i=1}^{N} X_{\stackidx{i}{i_2}{i}{j_2}},\quad \quad
\left(  \mathrm {tr}_B X\right)_{i_{1} j_{1}}  
:=\sum_{j=1}^{N} X_{\stackidx{i_1}{j}{j_1}{j}} 
\label{partrac}
\end{equation}

By the formula in Collins and {\'S}niady \cite{CS06} we have
\begin{eqnarray}
\label{eqn:int-formula-1-idx} 
\int_{\mathrm{U}\left(  N\right)} u_{i j} \overline{u_{i^{\prime} j^{\prime}}} dm\left(u\right)  
&=&\frac{1}{N} \delta_{i i^{\prime}}  \delta_{j j^{\prime}},
\\ \label{eqn:int-formula-2-idx}
\int_{\mathrm{U} \left(  N\right)} u_{i_{1} j_{1}}u_{i_{2} j_{2}}\overline
{u_{i_{1}^{\prime} j_{1}^{\prime}}u_{i_{2}^{\prime} j_{2}^{\prime}}}dm\left(u\right)  
&=&
\frac{1}{N^{2}-1}\left\{  
	\delta_{i_{1} i_{1}^{\prime}}
	\delta_{i_{2} i_{2}^{\prime}}  
	\delta_{j_{1} j_{1}^{\prime}}
	\delta_{j_{2} j_{2}^{\prime}}
+   \delta_{i_{1} i_{2}^{\prime}}
	\delta_{i_{2} i_{1}^{\prime}}
    \delta_{j_{1} j_{2}^{\prime}}
    \delta_{j_{2} j_{1}^{\prime}}
\right\}  
\\ \nonumber
&& 
-\frac{1}{N\left( N^{2}-1\right)  }
\left\{  
	\delta_{i_{1} i_{1}^{\prime}}
	\delta_{i_{2} i_{2}^{\prime}}  
	\delta_{j_{1} j_{2}^{\prime}}
	\delta_{j_{2} j_{1}^{\prime}}
+   \delta_{i_{1} i_{2}^{\prime}}
	\delta_{i_{2} i_{1}^{\prime}}
    \delta_{j_{1} j_{1}^{\prime}}
    \delta_{j_{2} j_{2}^{\prime}}
\right\} ,
\end{eqnarray}

where $dm$ denotes the Haar measure on $\mathrm{U}\left(  N\right)$. 

Let 
\begin{equation}
 z = \bra{x} U \otimes V X U^{\dagger} \otimes V^{\dagger} \ket{x}
\end{equation}
where $U,V$ are stochastically independent random unitary matrices of size $N$, 
distributed with Haar measure and $\ket{x}$ is an arbitrary vector. We are
interested in obtaining mean and variance of $z$, thus we will calculate  
$\mathbb{E} (z)$ and $\mathbb{E} (z \overline{z})$. We have
\begin{equation}
\mathbb{E}(z) = \mathbb{E}(\bra{x} U \otimes V X U^{\dagger} \otimes V^{\dagger} \ket{x}) 
      = \bra{x} \mathbb{E}(U \otimes V X U^{\dagger} \otimes V^{\dagger}) \ket{x} 
\end{equation}
and
\begin{eqnarray}
\!\!\!\!\!\!\!\!\!\!\!\!\mathbb{E}(z\overline{z}) &=& 
  \mathbb{E}(\bra{x} \otimes  \bra{\overline{x}} 
   (U \otimes V X U^{\dagger} \otimes V^{\dagger}) \otimes (\overline{U} \otimes 
   \overline{V} \, \overline{X} \, U^{T} \otimes V^{T})
   \ket{x} \otimes \ket{\overline{x}} )\\
 &=&
  \bra{x} \otimes  \bra{\overline{x}} 
  \ \mathbb{E}( (U \otimes V X U^{\dagger} \otimes V^{\dagger}) \otimes 
  (\overline{U} \otimes \overline{V} \, \overline{X} \, U^{T} \otimes V^{T})) \ 
  \ket{x} \otimes \ket{\overline{x}}.
\end{eqnarray}

To obtain the mean we calculate
\begin{eqnarray}
 \mathbb{E}(U \otimes V X U^{\dagger} \otimes V^{\dagger})  &=& 
 \left\{ \sum_{k,l=1}^{N^2} \mathbb{E} \left( (U \otimes V)_{i k} X_{k l} \overline{(U \otimes V)}_{j l} \right) \right\}_{i j} 
 \\
 &=& 
  \left\{ \sum_{k_1, k_2,l_1,l_2=1}^{N} \mathbb{E} (u_{i_1 k_1} v_{i_2 k_2} 
  X_{\stackidx{k_1}{k_2}{l_1}{l_2}}
 \overline{u}_{j_1 l_1} \overline{v}_{j_2 l_2} )   \right\}_{i j} 
 \\
 &=&
  \left\{ \sum_{k_1, k_2,l_1,l_2=1}^{N} \frac{1}{N}\frac{1}{N} \delta_{i_1 j_1} \delta_{k_1 l_1} \delta_{i_2 j_2} \delta_{k_2 l_2} 
  X_{\stackidx{k_1}{k_2}{l_1}{l_2}} 
  \right\}_{i j} 
 \\
 &=&
  \frac{1}{N^2} \tr X \ \mathrm{1}_{N^2 \times N^2},
\end{eqnarray}
where we have used a convention, that we split indices $\xi = N(\xi_1-1)+ \xi_2$, and $\xi_1, \xi_2$
has values from $\{1,2, \dots N \}$.
From the above, we have that for any normalized $\ket{x}$ 
\begin{equation}
\mathbb{E}(z) =   \mathbb{E}(\bra{x} U \otimes V X U^{\dagger} \otimes V^{\dagger} \ket{x})  = \frac{1}{N^2} \tr X.
\end{equation}
Let us now we calculate the second moment. Let  
\begin{equation}
M = \mathbb{E}( (U \otimes V X U^{\dagger} \otimes V^{\dagger}) \otimes (\overline{U} \otimes \overline{V} \, \overline{X} U^{T} \otimes V^{T})) 
\end{equation}
We have 
\begin{eqnarray}
&&\bra{i_1,i_2, k_1, k_2} M \ket{j_1,j_2, l_1, l_2} = \\
 && =  
 \sum_{\alpha, \beta, \gamma, \epsilon = 1}^{N^2}    
 \mathbb{E}( u_{i_1 \alpha_1} v_{i_2 \alpha_2} 
 X_{\stackidx{\alpha_1}{\alpha_2}{\beta_1}{\beta_2}}
  \overline{u}_{j_1 \beta_1} \overline{v}_{j_2 \beta_2}
  \overline{u}_{k_1 \gamma_1} \overline{v}_{k_2 \gamma_2} 
  \overline{X}_{\stackidx{\gamma_1}{\gamma_2}{\epsilon_1}{\epsilon_2}}
    u_{l_1 \epsilon_1} v_{l_2 \epsilon_2}
 ) \\
 && = 
 \sum_{\alpha, \beta, \gamma, \epsilon = 1}^{N^2}    
 \mathbb{E}( u_{i_1 \alpha_1} u_{l_1 \epsilon_1} \overline{u}_{j_1 \beta_1} \overline{u}_{k_1 \gamma_1} )
 \mathbb{E}( v_{i_2 \alpha_2} v_{l_2 \epsilon_2} \overline{v}_{j_2 \beta_2} \overline{v}_{k_2 \gamma_2} )
 X_{\stackidx{\alpha_1}{\alpha_2}{\beta_1}{\beta_2}}
 \overline{X}_{\stackidx{\gamma_1}{\gamma_2}{\epsilon_1}{\epsilon_2}},
\end{eqnarray} 
Using formula (\ref{eqn:int-formula-2-idx}) for the integral we have that 
$M = c_1 M^{(1)} + c_2 M^{(2)} + c_3 M^{(3)} + c_4 M^{(4)}$, where

\begin{eqnarray}
M^{(1)} &=&  \sum_{i_1,i_2,k_1,k_2 =1}^N\ket{i_1,i_2, k_1, k_2}  \bra{i_1,i_2, k_1, k_2}, 
\\ \nonumber 
c_1 &=&\frac{1}{(N^2-1)^2} \sum_{\alpha, \beta, \gamma, \epsilon = 1}^{N^2}   
\left(
 \delta_{\alpha_1 \beta_1} \delta_{\epsilon_1 \gamma_1} 
 -\frac{1}{N} \delta_{\alpha_1 \gamma_1} \delta_{\beta_1 \epsilon_1}
\right) \\
&&\left(
 \delta_{\alpha_2 \beta_2} \delta_{\epsilon_2 \gamma_2} 
 -\frac{1}{N} \delta_{\alpha_2 \gamma_2} \delta_{\beta_2 \epsilon_2}
\right)
 X_{\stackidx{\alpha_1}{\alpha_2}{\beta_1}{\beta_2}} 
 \overline{X}_{\stackidx{\gamma_1}{\gamma_2}{\epsilon_1}{\epsilon_2}}  
\\ \nonumber
&=& \frac{1}{(N^2-1)^2}
\Bigg(
 \sum_{\alpha_1,\alpha_2,\gamma_1, \gamma_2 = 1}^{N}    X_{\stackidx{\alpha_1}{\alpha_2}{\alpha_1}{\alpha_2}} 
 \overline{X}_{\stackidx{\gamma_1}{\gamma_2}{\gamma_1}{\gamma_2}}  
 -
 \frac{1}{N}\sum_{\alpha_1, \alpha_2 , \beta_2, \gamma_1 =1}^N  X_{\stackidx{\alpha_1}{\alpha_2}{\alpha_1}{\beta_2}} 
  \overline{X}_{\stackidx{\gamma_1}{\alpha_2}{\gamma_1}{\beta_2}}  \\
\nonumber &&\phantom{\frac{1}{(N^2-1)^2}} 
 -
 \frac{1}{N}\sum_{\alpha_1, \alpha_2 , \beta_1, \gamma_2 =1}^N  X_{\stackidx{\alpha_1}{\alpha_2}{\beta_1}{\alpha_2}} 
 \overline{X}_{\stackidx{\alpha_1}{\gamma_2}{\beta_1}{\gamma_2}}  
 + 
 \frac{1}{N^2}
 \sum_{\alpha_1, \alpha_2 , \beta_1, \beta_2 =1}^N  X_{\stackidx{\alpha_1}{\alpha_2}{\beta_1}{\beta_2}} 
  \overline{X}_{\stackidx{\alpha_1}{\alpha_2}{\beta_1}{\beta_2}}
 \Bigg) 
\\ \nonumber 
&=&\frac{1}{(N^2-1)^2}
\Bigg(
                \tr(X) \tr(X^{\dagger})
- \frac{1}{N}   \tr( \tr_A (X) \tr_A (X)^{\dagger})
- \frac{1}{N}   \tr( \tr_B (X) \tr_B (X)^{\dagger})\\&&
+ \frac{1}{N^2} \tr(X X^{\dagger})
\Bigg)
\\ \nonumber 
&=&\frac{1}{(N^2-1)^2}
\Bigg(
                |\tr X|^2
- \frac{1}{N}   \|\tr_A (X)\|_{\mathrm{HS}}^2
- \frac{1}{N}   \|\tr_B (X)\|_{\mathrm{HS}}^2 
+ \frac{1}{N^2} \|X\|_{\mathrm{HS}}^2 
\Bigg).
\end{eqnarray}
Similarly 
\begin{eqnarray}
M^{(2)} &=& \sum_{i_1,i_2,j_1,k_2=1}^N \ket{j_1,i_2, j_1,k_2}\bra{i_1,i_2,i_1,k_2}, 
\\ \nonumber 
c_2 &=& \frac{1}{(N^2-1)^2} \sum_{\alpha, \beta, \gamma, \epsilon = 1}^{N^2}   
\left(
  \delta_{\alpha_1 \gamma_1} \delta_{\beta_1 \epsilon_1}  
 -\frac{1}{N} \delta_{\alpha_1 \beta_1} \delta_{\epsilon_1 \gamma_1} 
\right) \\&&
\left(
 \delta_{\alpha_2 \beta_2} \delta_{\epsilon_2 \gamma_2} 
 -\frac{1}{N} \delta_{\alpha_2,\gamma_2} \delta_{\beta_2, \epsilon_2}
\right)
 X_{\stackidx{\alpha_1}{\alpha_2}{\beta_1}{\beta_2}} 
 \overline{X}_{\stackidx{\gamma_1}{\gamma_2}{\epsilon_1}{\epsilon_2}}  
\\ \nonumber 
&=&\frac{1}{(N^2-1)^2}
\left(
  \|\tr_B (X)\|_{\mathrm{HS}}^2 
- \frac{1}{N} \|X\|_{\mathrm{HS}}^2 
- \frac{1}{N} |\tr X|^2
+ \frac{1}{N^2} \|\tr_A (X)\|_{\mathrm{HS}}^2
\right).
\end{eqnarray}
Next we have 
\begin{eqnarray}
&&
M^{(3)} = \sum_{i_1,i_2,j_2,k_1=1}^N \ket{i_1,j_2, k_1,j_2}\bra{i_1,i_2,k_1,i_2},  
\\
\nonumber &&
c_3=\frac{1}{(N^2-1)^2} 
\left(
                \|\tr_A (X)\|_{\mathrm{HS}}^2 
- \frac{1}{N}   \|X\|_{\mathrm{HS}}^2 
- \frac{1}{N}    |\tr X|^2
+ \frac{1}{N^2} \|\tr_B (X)\|_{\mathrm{HS}}^2 
\right).
\end{eqnarray}
Finally 
\begin{eqnarray}
&&
M^{(4)} = \sum_{i_1,i_2,j_i,j_2=1}^N\ket{j_1,j_2, j_1,j_2}\bra{i_1,i_2,i_1,i_2},\\ 
\nonumber  &&
c_4=\frac{1}{(N^2-1)^2}
\left(
                \|\tr_B (X)\|_{\mathrm{HS}}^2 
- \frac{1}{N}  \|\tr_A (X)\|_{\mathrm{HS}}^2 
- \frac{1}{N}   \|\tr_B (X)\|_{\mathrm{HS}}^2 
+ \frac{1}{N^2}  |\tr X|^2
\right).
\end{eqnarray}

The integrals of the above type, for fixed dimension, can be calculated with 
the use of computer algebra program \texttt{IntU}~\cite{PM11int}.

\section{Variance values for shadows with fixed Schmidt numbers}
Let us consider any pure state $\ket{x}$ from $N \times N$ composite Hilbert
space ${\cal H}={\cal H}_A \otimes {\cal H}_B$ with  fixed Schmidt numbers
$\lambda_1, \lambda_2, \dots , \lambda_N$, thus we have  
\begin{equation} 
\ket{x} = \sum_{i=1}^N \sqrt{\lambda_i} \ket{i^A} \otimes\ket{i^B}, 
\end{equation} 
for some orthogonal bases $\ket{i^A},\ket{i^B}$. For simplicity we can take computational
bases. We have the following identities 
\begin{eqnarray}
\bra{x}\otimes \bra{\overline{x}} M_1 \ket{x} \otimes \ket{\overline{x}} &=& \bra{x}\otimes \bra{\overline{x}} M_4 \ket{x} \otimes \ket{\overline{x}}= 1 \\
\bra{x}\otimes \bra{\overline{x}} M_2 \ket{x} \otimes \ket{\overline{x}} &=& \bra{x}\otimes \bra{\overline{x}} M_3 \ket{x} \otimes \ket{\overline{x}} = \sum_{i=1}^N \lambda_i^2.
\end{eqnarray}
Thus we have,
\begin{eqnarray}
\bra{x}\otimes \bra{\overline{x}} M \ket{x} \otimes \ket{\overline{x}} &=& 
 c_1 + c_4 + (c_2 + c_3)\sum_{i=1}^N \lambda_i^2,
\end{eqnarray}
Note that, the above depends only on a purity of reduced state. In the case of
maximally entangled state ($\lambda_i  = \frac{1}{N}$ for $i=1,2,\dots N$) the
second moment is given by 
\begin{eqnarray}
&\bra{x}\otimes \bra{\overline{x}} M \ket{x} \otimes \ket{\overline{x}} = &
\frac{1}{N^2(N^2-1)} 
\left( 
  |\tr X|^2 
  +  
 \|X\|_{\mathrm{HS}}^2 
\right)\\&&
-\frac{1}{N^3(N^2-1)}  \left(
 \|\tr_A (X)\|_{\mathrm{HS}}^2  
 + 
 \|\tr_B (X)\|_{\mathrm{HS}}^2 
\right) .
\end{eqnarray}
The above implies the result (\ref{distestshad}) for the variance with respect to
the entangled shadow.

In the case of separable states ($\lambda_i  = \delta_{1,i}$ for $i=1,2,\dots
N$) we obtain 
\begin{eqnarray}
&&\bra{x}\otimes \bra{\overline{x}} M \ket{x} \otimes \ket{\overline{x}} = \\
&&\frac{1}{N^2 (N+1)^2} 
 \left(
    |\tr X|^2 
+  \|\tr_A (X)\|_{\mathrm{HS}}^2  
+  \|\tr_B (X)\|_{\mathrm{HS}}^2 
+  \|X\|_{\mathrm{HS}}^2 
\right).
\end{eqnarray}
Which implies the result (\ref{var-sep}) for the variance with respect to the
separable shadow.


\end{document}